\documentstyle[here,epsfig]{mn2e}
\def\simlt{\mathrel{\rlap{\lower 3pt\hbox{$\sim$}}\raise 2.0pt\hbox{$<$}}}
\def\simgt{\mathrel{\rlap{\lower 3pt\hbox{$\sim$}} \raise 2.0pt\hbox{$>$}}}

\def\gtsima{$\; \buildrel > \over \sim \;$}
\def\ltsima{$\; \buildrel < \over \sim \;$}
\def\gtrsim{\lower.5ex\hbox{\gtsima}}
\def\lesssim{\lower.5ex\hbox{\ltsima}}
\def\url#1{{\ttfamily\def\/{/\diskretionary{}{}{}}#1}}
\begin{document}

\newcommand{\q}{\begin{equation}}
\newcommand{\qa}{\begin{eqnarray}}
\newcommand{\qs}{\begin{eqnarray*}}
\newcommand{\nq}{\end{equation}}
\newcommand{\nqa}{\end{eqnarray}}
\newcommand{\nqs}{\end{eqnarray*}}
\newcommand{\ud}{\mathrm{d}}

\title[IMBHs and ULXs in Cartwheel] 
{Intermediate-mass black holes and ultraluminous X-ray sources in the Cartwheel ring galaxy}
\author[Mapelli et al.] 
{M. Mapelli$^{1}$, B. Moore$^{1}$, L. Giordano$^{2}$, L. Mayer$^{1,3}$, M. Colpi$^{2}$, E. Ripamonti$^{2,4}$,
  \newauthor S. Callegari$^{1}$ 
\\ 
$^{1}$ Institute for Theoretical Physics, University of Z\"urich, Winterthurerstrasse 190, CH-8057, Z\"urich, Switzerland; {\tt mapelli@physik.unizh.ch}\\
$^{2}$ Universit\`a Milano Bicocca, Dipartimento di Fisica G.~Occhialini, Piazza delle Scienze 3, I-20126, Milano, Italy\\
$^{3}$ Institute of Astronomy, ETH Z\"urich, ETH Honggerberg HPF D6, CH-8093, Z\"urich, Switzerland\\
$^{4}$ Kapteyn Astronomical Institute, University of Groningen, Postbus 800, 9700 AV Groningen, the Netherlands\\
}

\maketitle \vspace {7cm }

\begin{abstract}
$Chandra$ and $XMM-Newton$ observations of the Cartwheel galaxy show $\sim{}17$ bright X-ray sources ($\gtrsim{}5\times{}10^{38}$ erg s$^{-1}$), all within the gas-rich outer ring. We explore the hypothesis that these X-ray sources are powered by intermediate-mass black holes (IMBHs) accreting gas or undergoing mass transfer from a stellar companion. To this purpose, we run $N$-body/SPH simulations of the galaxy interaction which might have led to the formation of Cartwheel, tracking the dynamical evolution of two different IMBH populations: halo and disc IMBHs. Halo IMBHs cannot account for the observed X-ray sources, as only a few of them cross the outer ring. Instead, more than half of the disc IMBHs are pulled in the outer ring as a consequence of the galaxy collision.
However, also in the case of disc IMBHs, accretion from surrounding gas clouds cannot account for the high luminosities of the observed sources. 
Finally, more than 500 disc IMBHs are required to produce $\lesssim{}15$ X-ray sources via mass transfer from very young stellar companions. Such number of IMBHs is very large and implies extreme assumptions. Thus, the hypothesis that all the observed X-ray sources in Cartwheel are associated with IMBHs is hardly consistent with our simulations, even if it is still possible that IMBHs account for the few ($\lesssim{}1-5$) brightest ultraluminous X-ray sources (ULXs).
\end{abstract}

\section{Introduction}
Ring galaxies are among the most fascinating objects in the Universe. The Cartwheel galaxy is surely the most famous of them, and also the biggest (with its optical diameter of $\sim{}$50-60 kpc) and the most studied. Cartwheel has been thoroughly observed in almost every band: H$\alpha{}$ (Higdon 1995) and optical (Theys \& Spiegel 1976, Fosbury \& Hawarden 1977) images, red continuum (Higdon 1995), radio line (Higdon 1996) and continuum (Higdon 1996; Mayya et al. 2005), near- (Marcum, Appleton \& Higdon 1992) and far-infrared (Appleton \& Struck-Marcell 1987a), line spectroscopy (Fosbury \& Hawarden 1977) and X-ray (Wolter, Trinchieri \&{} Iovino 1999; Gao et al. 2003; Wolter \& Trinchieri 2004; Wolter, Trinchieri \& Colpi 2006). 


Cartwheel exhibits a double ringed shape, with some transversal 'spokes' (Theys \& Spiegel 1976, Fosbury \& Hawarden 1977; Higdon 1995), which have been detected only in few of the $\lesssim{}300$ known ring galaxies (Arp \& Madore 1987; Higdon 1996). The outer ring is rich of HII regions, especially in its southern quadrant, and dominates the H$\alpha{}$ emission (Higdon 1995). This implies that Cartwheel is currently  undergoing an intense epoch of star formation (SF, with SF rate $\sim{}20-30\,{}M_\odot{}$ yr$^{-1}$; Marston \& Appleton 1995; Mayya et al. 2005), almost entirely confined to the outer ring. Inner ring, nucleus and spokes are nearly devoid of gas and dominated by red continuum emission, indicating a  relatively old, low- and intermediate-mass stellar population (Higdon 1995, 1996; Mayya et al. 2005). 
Cartwheel is located in a small group of 4 galaxies. All of its 3 companions (known as G1, G2 and G3) are less massive than Cartwheel, and host less than 20 per cent of the total gas mass in the group.

The analysis of X-ray data shows another intriguing peculiarity: most of the point sources detected by $Chandra$ are in the outer ring, and particularly concentrated in the southern quadrant (Gao et al. 2003; Wolter \& Trinchieri 2004). According to Wolter \& Trinchieri (2004) 13 out of 17 sources associated with Cartwheel are in the outer ring, the remaining 4 being close to the inner rim of the ring or to the optical spokes\footnote{The total number of point X-ray sources in the Cartwheel group is 24 (Wolter \& Trinchieri 2004), but 6 of them are associated with the companion galaxies G1 and G2, whereas 1 of them is probably background/foreground contamination.}. Gao et al. (2003) noted that all the five strongest H$\alpha{}$ knots in the ring are associated with an X-ray source, indicating a possible correlation between X-ray sources and young star-forming regions. Furthermore, most of the observed sources have isotropic X-ray luminosity $L_X\gtrsim{}10^{39}$ erg s$^{-1}$, belonging to the category of ultraluminous X-ray sources (ULXs). Given the distance of Cartwheel ($\sim{}124$ Mpc for an Hubble constant $H_0=73$ km s$^{-1}$ Mpc$^{-1}$), the data are incomplete for $L_X\lesssim{}5\times{}10^{38}$ erg s$^{-1}$, and the faintest sources in the sample have $L_X\sim{}10^{38}$ erg s$^{-1}$. Then, almost all the X-ray sources detected in Cartwheel are close to the ULX range.

Many theoretical studies, both analytical (Lynds \& Toomre 1976; Theys \& Spiegel 1976; Appleton \& Struck-Marcell 1996) and numerical (Theys \& Spiegel 1976; Appleton \& Struck-Marcell 1987a, 1987b; Hernquist \& Weil 1993; Mihos \& Hernquist 1994; Struck 1997; Horellou \& Combes 2001; Griv 2005) were aimed to explain the origin of Cartwheel and its observational features. In the light of these papers, the origin of propagating rings in Cartwheel and similar galaxies can be explained by galaxy collisions with small impact parameter (Theys \& Spiegel 1976; Appleton \& Struck-Marcell 1987a, 1987b; Hernquist \& Weil 1993; Mihos \& Hernquist 1994; Struck 1997; Horellou \& Combes 2001). Among the companions of Cartwheel, both G1 and G3 are good candidates for this interaction, having a small projected impact parameter, being not too far from Cartwheel, and showing a disturbed distribution of neutral hydrogen (Higdon 1996).
An alternative, less investigated model explains the rings with disc instabilities (Griv 2005). 

Models of galaxy collisions explain quite well most of Cartwheel properties.
However, none of the previous theoretical studies has investigated the nature of the X-ray sources, and especially of the ULXs, observed in Cartwheel.

The nature of the ULXs is still not understood. It has been suggested that they are  high-mass X-ray binaries (HMXBs) powered by  stellar mass black holes (BHs) with anisotropic X-ray emission (King et al. 2001; Grimm, Gilfanov \& Sunyaev 2003; King 2006)  or with super-Eddington accretion rate (e.g. King \&{} Pounds 2003; Socrates \& Davis 2006; Poutanen et al. 2007). However, some ULXs, especially the brightest ones ($L_X\gtrsim{}10^{40}$ erg s$^{-1}$), show characteristics which are difficult to reconcile with  the hypothesis of beamed emission, such as the presence of an isotropically ionized nebula (e.g. Pakull \& Mirioni 2003; Kaaret, Ward \& Zezas 2004) or of quasi periodic oscillations (e.g. Strohmayer \& Mushotzky 2003). Then, it has been proposed that  some ULXs (or at least the brightest among them; Miller, Fabian \& Miller 2004) might be associated with intermediate-mass black holes (IMBHs), i.e. BHs with mass $20\lesssim{}m_{\rm BH}/M_\odot{}\lesssim{}10^{5}$, accreting either gas or companion stars (Miller et al. 2004; see Mushotzky 2004 and Colbert \& Miller 2005 for a review). On the other hand, most of ULXs can be explained with the properties of stellar mass  BHs (see Roberts 2007 for a review and references therein). 

In this paper we present a new, refined $N-$body/SPH model of the Cartwheel galaxy, which reproduces the main features of Cartwheel.
 The aim of this paper is to use the $N-$body/SPH model in order to check whether the IMBH  hypothesis is viable to explain all or a part of the X-ray sources observed in Cartwheel. In fact, in the last few years the hypothesis that most of the ULXs are powered by  IMBHs has became increasingly difficult to support, as the observational features of the majority of ULXs are consistent with accreting stellar mass BHs (Roberts 2007). It would be interesting to see whether also the dynamical simulations agree with this conclusion\footnote{We will not consider other possible scenarios (e.g. the production of ULXs by beamed emission in HMXBs), due to intrinsic limits of $N-$body methods.}.

In Section 2 we describe our simulations. In Section 3 we discuss the evolution of our models and the dynamics of either halo or disc IMBHs. In Section 4 we investigate the possibility that the X-ray sources in Cartwheel are powered by IMBHs accreting gas or stars. In the last case we consider both the accretion from old stars (which produce only transient sources) and from young stars, formed after the galaxy collision (which can lead also to persistent sources). We also investigate the hypothesis that new disc IMBHs are formed in very young clusters after the galaxy collision. Our findings are summarized in Section 5.

\section{The $N$-body model}
In order to form a 'Cartwheel-like' galaxy, we simulate the encounter between a disc galaxy (in the following, we will call it 'progenitor' of Cartwheel) and a smaller mass companion (in the following 'intruder'), integrating the evolution of the system over $\sim{}1$ Gyr.

The simulations have been carried out running the parallel $N$-body/SPH code GASOLINE (Wadsley, Stadel \& Quinn 2004) on the cluster $zBox2$\footnote{\tt http://www-theorie.physik.unizh.ch/$\sim{}$dpotter/zbox/} at the University of Z\"urich.

\subsection{The progenitor of Cartwheel}
The $N$-body model adopted to simulate the 'progenitor' of the Cartwheel galaxy is analogous to that described in Mapelli, Ferrara \& Rea (2006) and in Mapelli (2007), derived from the recipes by Hernquist (1993). The progenitor galaxy is represented by four different components: a Navarro, Frenk \& White (1996, hereafter NFW) halo, an Hernquist bulge (absent in some runs), an exponential stellar disc and an exponential gaseous disc. Halo, bulge, disc and gas velocities are generated  using the Gaussian approximation (Hernquist 1993), as described in Mapelli et al. (2006).
\subsubsection{The halo}
We adopt a NFW halo, according to the formula (NFW; Moore et al. 1999)
\q\label{eq:eq1}
\rho_h(r)=\frac{\rho_s}{(r/r_s)^\gamma{}\,{}[1+(r/r_s)^\alpha{}]^{(\beta{}-\gamma{})/\alpha{}}},
\nq
where we choose $(\alpha{},\beta{}, \gamma{})=(1,3,1)$, and
$\rho{}_s=\rho_{crit}\,{}\delta_c$, $\rho_{crit}$ being the critical
density of the Universe and
\begin{equation}\label{eq:eq2}
\delta_c=\frac{200}{3}\frac{c^3}{\ln{(1+c)}-[c/(1+c)]},  
\end{equation}
where $c$ is the concentration parameter and $r_s$ is the halo scale radius, defined by
$r_s=R_{200}/c$. $R_{200}$ is the radius encompassing a mean overdensity of 200 with respect to the 
background density of the Universe.
  $R_{200}$ can be calculated as $R_{200}=V_{200}/[10\,{}H(z)]$, where $V_{200}$ is the circular velocity at the
virial radius and $H(z)$ is the Hubble parameter at the redshift $z$. 

As the measured mass of Cartwheel within the virial radius is $\sim{}3-6\times{}10^{11}M_\odot{}$ (Higdon 1996; Horellou \& Combes 2001), we adopt $V_{200}=100$ km s$^{-1}$, obtaining $R_{200}=140$ kpc. Adopting a fiducial concentration $c=12$, we obtain $r_s=12$ kpc.

We made check simulations changing some of those parameters (e.g. with $c=5, 10$) without observing any significant difference in the post-encounter evolution.

\subsubsection{The bulge}
We adopt a spherical Hernquist bulge
\q\label{eq:eq4}
\rho_b(r)=\frac{M_b\,{}a}{2\pi}\,{}\frac{1}{r\,{}(a+r)^3},
\nq
where $M_b$ and $a$ are  the bulge mass and scale length (see Section 2.4 and Table 1). 

\subsubsection{The disc}
The stellar disc profile is (Hernquist 1993):
\q\label{eq:eq3}
\rho_d(R,z)=\frac{M_d}{4\pi R_d^2\,{}z_0}\,{}e^{-R/R_d}\,{}\textrm{sech}^2(z/z_0),
\nq
where $M_d$ is the disc mass, $R_d$ the disc scale length and $z_0$ the disc scale height. We adopt different values of these parameters for different runs (see  Section 2.4 and Table 1).

\subsubsection{The gaseous disc}
 For the gas component we adopt the same exponential profile as in equation~(\ref{eq:eq3}), with different values for the total mass $M_g$, the scale length $R_g$ and scale height $z_g$ (see Section 2.4 and Table 1).
The gas is allowed to cool down to a temperature of $2\times{}10^4$ K. In some runs we also switch-on SF, according to the Schmidt law (Katz 1992; Wadsley et al. 2004).

\subsection{The intruder}
We model the intruder as a NFW halo with total mass $3.2\times{}10^{11}\,{}M_\odot{}$, i.e. a fraction of $\sim{}0.5-0.6$ (depending on the run) of the total mass of the Cartwheel progenitor. We assume $R_{200}=30$ kpc and  $c=12$. Less concentrated intruders produce much less developed rings.
We made also runs where the intruder has an exponential gas disc of $2\times{}10^9\,{}M_\odot{}$ (consistent with the mass in gas of G3, one of the most probable intruders). 
 
We put the intruder on an orbit with a null (Hernquist \& Weil 1993) or small (Horellou \& Combes 2001) impact parameter (see Table 1), with a centre-of-mass velocity, relative to Cartwheel, close to the escape velocity.

\subsection{Intermediate-mass black holes}
Not only the properties, but the very existence of IMBHs are still uncertain (see van der Marel 2004 for a review). Then, we do not know anything from the observations about the positions, the velocities and the mass function of IMBHs in galaxies. 

Different theoretical models predict different properties for IMBHs. We will focus  on two of these models: (i) the formation of IMBHs from massive metal-free\footnote{As far as we know, IMBHs cannot be formed as remnants of recent (i.e. relatively high metallicity) SF. In fact, population I 
massive stars lose a significant fraction of their mass due to stellar winds before the
formation of the BH (Heger et al. 2003). On the contrary, BHs with mass $> 20\,{}M_\odot{}$ might have formed as remnants of population III stars, which, being metal-free, are not significantly affected 
by mass loss  (Heger \&{} Woosley 2002; Heger et al. 2003).} stars (Heger \& Woosley 2002); (ii) the runaway collapse of stars in young clusters (Portegies Zwart \& McMillan 2002).
In the former scenario, IMBHs are the remnants of very massive ($30-140\,{}M_\odot{}$ or $\gtrsim{}260\,{}M_\odot{}$, Heger \& Woosley 2002) population III stars. Then, they are expected to form at high redshift ($\approx{}10-25$) in mini-haloes. Diemand, Madau \& Moore (2005, hereafter DMM) showed that objects formed in high density peaks at high redshift appear more concentrated than more recent structures. According to DMM, the density profile of objects formed in a $\nu{}\,{}\sigma{}$ peak can be parametrized as
\q\label{eq:DMM}
\rho{}_{\rm BH}(r)=\frac{\rho_s}{(r/r_\nu)^\gamma{}\,{}[1+(r/r_\nu{})^\alpha{}]^{(\beta{}_\nu{}-\gamma{})/\alpha{}}},
\nq
where $\alpha{}$ and $\gamma{}$ are the same as defined in equation~(\ref{eq:eq1});  $r_\nu\equiv{}r_s/f_\nu{}$ is the scale radius for objects formed in a  $\nu{}\,{}\sigma{}$ fluctuation [with $f_\nu{}=\exp{(\nu{}/2)}$], and $\beta{}_\nu{}=3+0.26\,{}\nu{}^{1.6}$. Hereafter, we will refer to equation~(\ref{eq:DMM}) as the DMM profile. This profile is likely to track the distribution of IMBHs, if they are a halo population (for alternative models see e.g. Bertone, Zentner \& Silk 2005).

In the second scenario, repeated collisions in star clusters with sufficiently small initial half-mass relaxation time ($t_h\lesssim{}25$ Myr) lead to the runaway growth of a central object with mass up to 0.1 per cent of the total mass of the cluster (Portegies Zwart \& McMillan 2002). All clusters with current $t_h\lesssim{}100$ Myr are expected to form IMBHs in this way. As young clusters are a disc population, also the IMBHs formed via runaway collapse are expected to lie in the disc. As shown by Mapelli (2007) disc IMBHs can accrete more efficiently than halo IMBHs, leading to stronger constraints on their number and mass.

In this paper, we run simulations  where IMBHs are modelled either as a disc or as a halo population. As described in Mapelli (2007), halo IMBHs are distributed according to a DMM profile for objects formed in $3.5\,{}\sigma{}$ fluctuations; while disc IMBHs are distributed in the same way as stars (see equation~\ref{eq:eq3}).

 In the following, we will consider IMBHs of $10^2$ and $10^3\,{}M_\odot{}$, which are the typical masses suggested both by the first star scenario and by the runaway collapse.  Furthermore, such mass range has also been deeply studied from the point of view of the mass-transfer history (Portegies Zwart, Dewi \& Maccarone 2004; Blecha et al. 2006; Madhusudhan et al. 2006). In particular, we will consider IMBHs of $10^3\,{}M_\odot{}$ as an upper limit only in the case of gas accretion from surrounding clouds. In all the other cases we will assume that the IMBHs have a mass of $10^2\,{}M_\odot{}$ as our fiducial case. Larger masses of the IMBHs can be hardly reconciled with the observed properties of most of ULXs, as highlighted by Roberts (2007). In fact, both the shape of the X-ray luminosity function (Grimm et al. 2003) and the association of many ULXs with star-forming regions (King 2004) suggest that ULXs are dominated by BHs of mass up to $\sim{}100\,{}M_\odot{}$ (Roberts 2007). Also the analysis of the spectra of some bright ULXs seem to confirm this mass range, showing the intrinsic weaknesses of cool black-body disc models  (Goncalves \& Soria 2006; Stobbart, Roberts \&{} Wilms 2006). Finally, also the analysis of the optical counterparts of 7 ULXs shows that the host BHs are consistent with stellar mass BHs or with $\sim{}100\,{}M_\odot{}$ IMBHs (Copperwheat et al. 2007).

\subsection{Description of runs}
\begin{table}
\begin{center}
\caption{Initial parameters of runs
}
\begin{tabular}{lllll}
\hline
\vspace{0.1cm}
Run &  $M_d/(10^{10}M_\odot{})$ & $M_b/(10^{10}M_\odot{})$ & IMBH profile & SF \\ 
\hline
A1  &  9.6 & 2.4 & DMM & no \\
A2  &  9.6 & 2.4 & disc & no \\
\vspace{0.1cm}
A3  &  9.6 & 2.4 & disc & yes \\
B1  &  4.8 & 0 & DMM & no \\
B2  &  4.8 & 0 & disc & no \\
B3  &  4.8 & 0 & disc & yes \\
\hline
\end{tabular}
\end{center}
\label{tab_1}
\end{table}

In all the performed runs the progenitor galaxy has 122000 dark matter halo particles with a mass of $4\times{}10^6M_\odot{}$, corresponding to a total halo mass of 4.9$\times{}10^{11}M_\odot{}$, consistent with the observations (Higdon 1995). The intruder is always composed by 80000 dark matter particles, for a total mass of 3.2$\times{}10^{11}M_\odot{}$. 

Disc, bulge and gas particles, as well as the particles hosting an IMBH\footnote{Particles hosting an IMBH are normal stellar particles. We assume that a fraction of their mass represents an IMBH of 10$^2$ or $10^3$~$M_\odot{}$, while the remaining mass is composed by stars.}, have mass equal to $4\times{}10^5M_\odot{}$. The number of IMBH particles in each simulation is fixed to the reference value of 100 (but the results can be easily rescaled). The initial number of gas particles is equal to 80000 (corresponding to $M_g=3.2\times{}10^{10}M_\odot{}$) in all the simulations. The initial $M_g$ in our simulations is about a factor of 1.5 higher than the observed value (Higdon 1995), to allow a  fraction of initial gas to form stars (Cartwheel is a starburst galaxy) or be stripped during the interaction.

The number of disc and bulge particles depends on the simulation. As it is shown in Table~1, runs labelled as 'A' have $M_d=9.6\times{}10^{10}M_\odot{}$ and  $M_b=2.4\times{}10^{10}M_\odot{}$ (corresponding to 240000 and 60000 disc and bulge particles, respectively). Runs A have $R_d=4.4$ kpc, $z_0=0.1\,{}R_d$ and $a=0.2\,{}R_d$. The analogous properties for the gaseous disc in runs A are $R_g=R_d$ and $z_g=0.057\,{}R_g$ (where $R_g$ and $z_g$ are the gas disc scale length and scale height, see Section 2.1).

Runs labelled as 'B' in Table~1 have $M_d=4.8\times{}10^{10}M_\odot{}$ (corresponding to 120000 star particles), $M_b=0$, $R_d=6.6$ kpc, $z_0=0.2\,{}R_d$, $R_g=R_d$ and $z_g=0.057\,{}R_g$. The characteristics of runs B (and, in particular, the absence of a bulge) have been chosen in analogy with Hernquist \& Weil (1993); while the properties of runs A are similar to those used by Horellou \& Combes (2001). As we will see in the next section, the presence of the bulge does not affect significantly the results. Instead, the number of pre-existing stars in the disc can be important for the formation of spokes.

Softening lengths\footnote{For the choice of the halo softening we adopt the recipe by Dehnen (2001). For disc particles a (up to a factor of $\sim{}$20) larger softening than the adopted one gives essentially the same results. 
For gas particles we do not have fragmentation problems, as the Jeans mass is a factor of $>20$ larger than the mass resolution during the entire simulation (see Bate \& Burkert 1997).} are 0.2 kpc for dark matter particles and 0.01 kpc for disc, bulge, IMBH and gas particles. The initial smoothing length is also 0.01 kpc.

There is another fundamental difference between runs A and B, i.e. the inclination of the initial velocity direction of the intruder with respect to the symmetry axis of the Cartwheel progenitor. The initial position and velocity of the centre-of-mass of the intruder are {\bf x}=(0, 8, 32) kpc and {\bf v}=(28, -218, -872) km s$^{-1}$ for runs A, and {\bf x}=(0, 0, 40) kpc and {\bf v}=(0, 0, 900) km s$^{-1}$ for runs B. This implies that the intruder travels along the symmetry axis of Cartwheel in the runs B (as in Hernquist \& Weil 1993), while it is a bit off-centre in runs A (as in Horellou \& Combes 2001). As we will see in the next Section, this difference is crucial for the shape of the post-encounter Cartwheel.

Finally, in runs A1 and B1 the IMBHs are distributed according to a DMM profile for 3.5$\,{}\sigma{}$ fluctuations; whereas IMBHs populate the exponential stellar disc in runs A2, A3, B2 and B3. In runs A3 and B3 SF is allowed, with an efficiency $c_\ast{}=0.1$.


\section{Dynamical evolution of Cartwheel models}
\begin{figure}
\center{{
\epsfig{figure=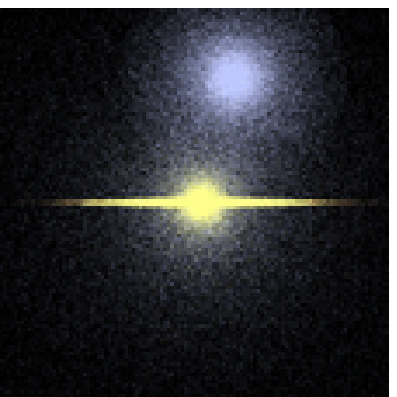,height=4cm}
\epsfig{figure=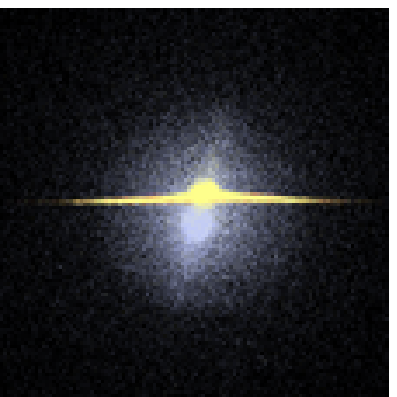,height=4cm}
\epsfig{figure=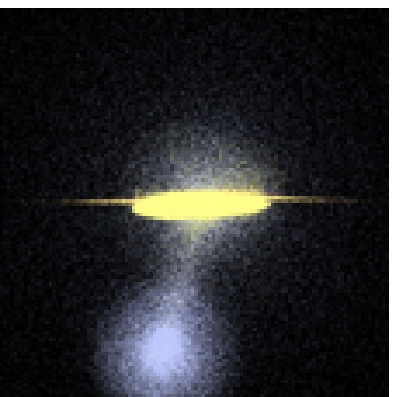,height=4cm}
\epsfig{figure=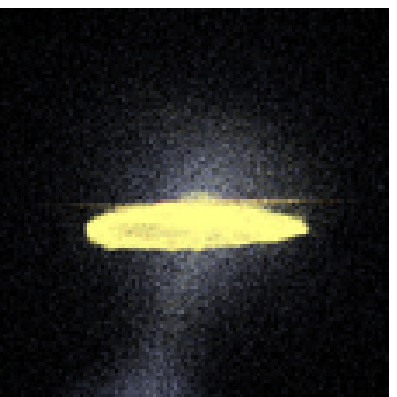,height=4cm}
\epsfig{figure=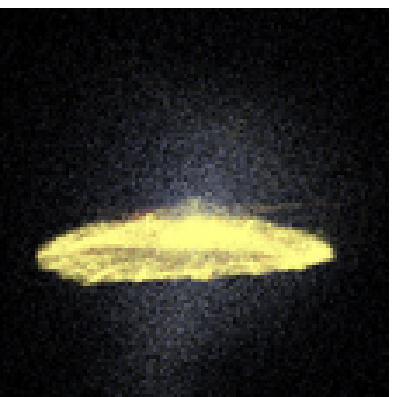,height=4cm}
\epsfig{figure=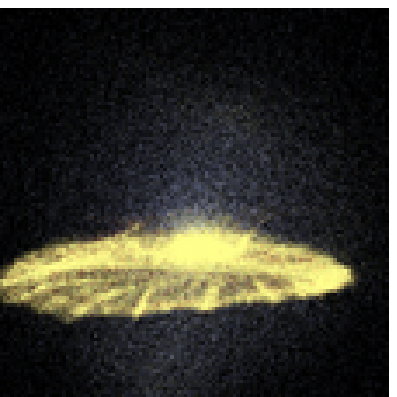,height=4cm}
}}
\caption{\label{fig:fig1} 
Run A3, time evolution of the dark matter (blue on the web), stellar (yellow on the web) and gas (red on the web) component during the encounter. From top to bottom and from left to right: after 0, 40, 80, 120, 160 and 200~Myr from the beginning of the simulation. Each frame measures 100 kpc per edge. Cartwheel is seen edge-on.} 
\end{figure}

\begin{figure}
\center{{
\epsfig{figure=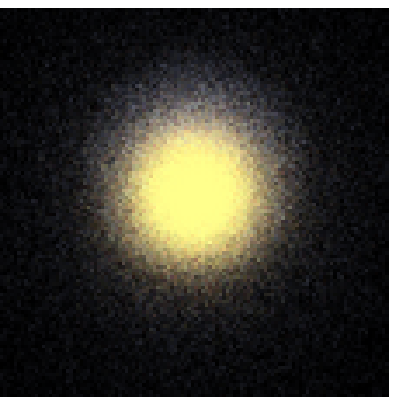,height=4cm}
\epsfig{figure=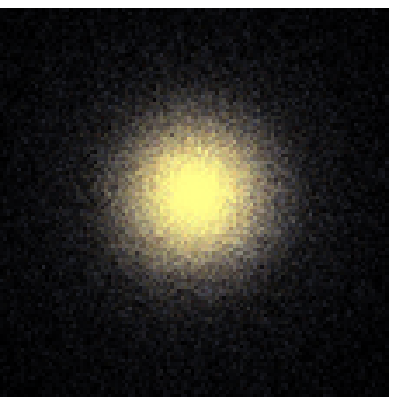,height=4cm}
\epsfig{figure=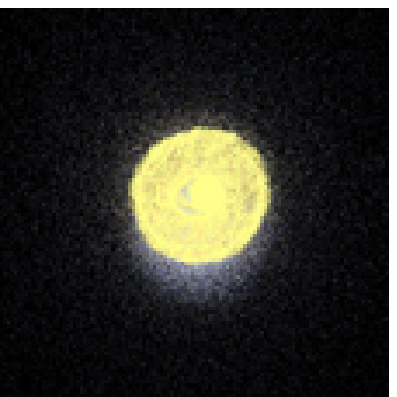,height=4cm}
\epsfig{figure=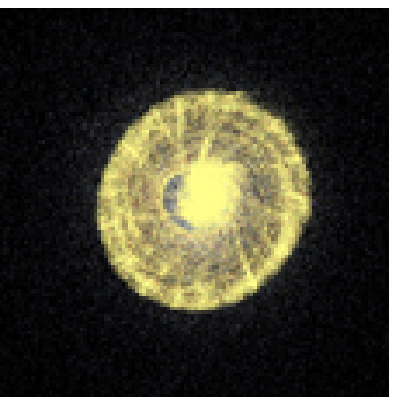,height=4cm}
\epsfig{figure=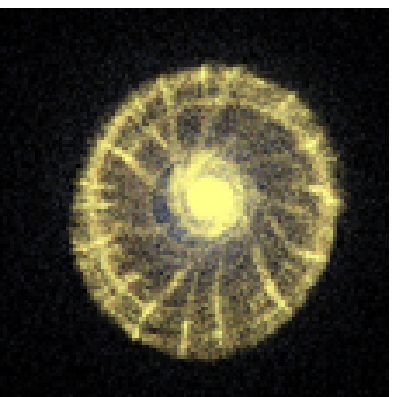,height=4cm}
\epsfig{figure=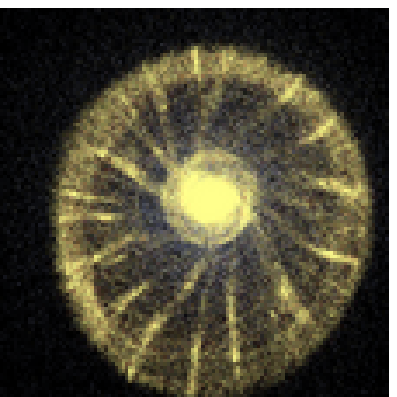,height=4cm}
}}
\caption{\label{fig:fig2} 
The same as Fig.~\ref{fig:fig1}, but Cartwheel is seen face-on.} 
\end{figure}

\begin{figure}
\center{{
\epsfig{figure=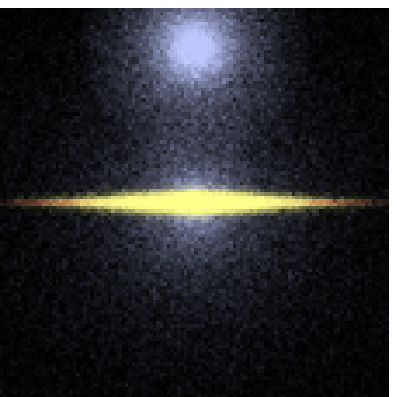,height=4cm}
\epsfig{figure=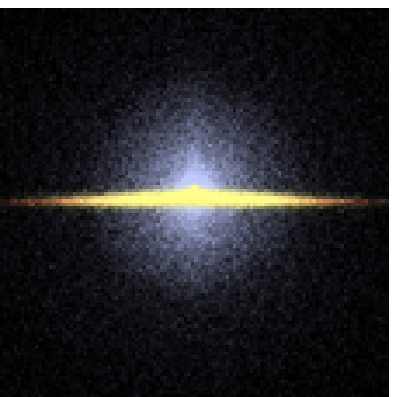,height=4cm}
\epsfig{figure=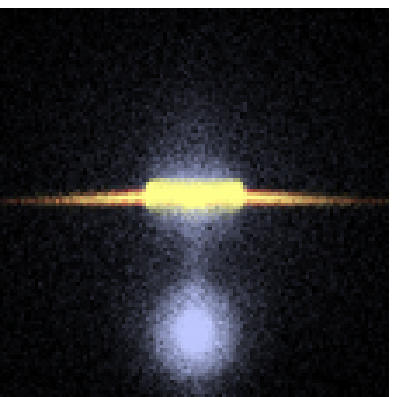,height=4cm}
\epsfig{figure=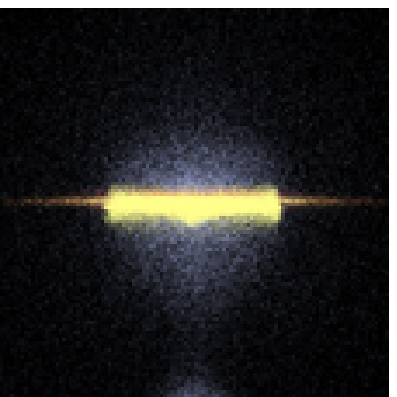,height=4cm}
\epsfig{figure=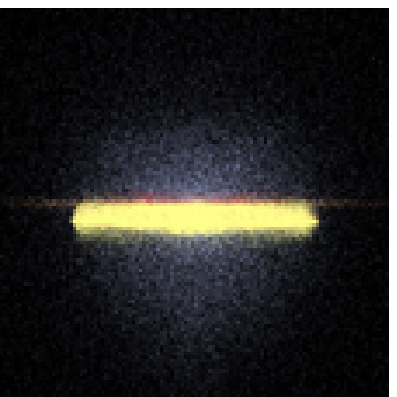,height=4cm}
\epsfig{figure=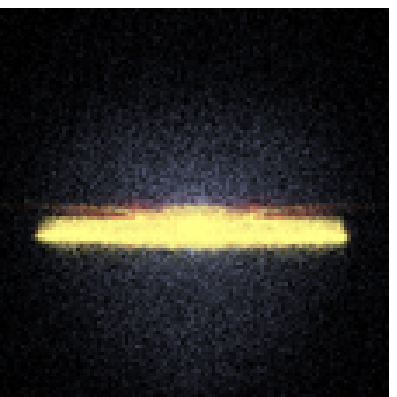,height=4cm}
}}
\caption{\label{fig:fig3} 
The same as Fig.~\ref{fig:fig1}, but for run B3.}
\end{figure}

\begin{figure}
\center{{
\epsfig{figure=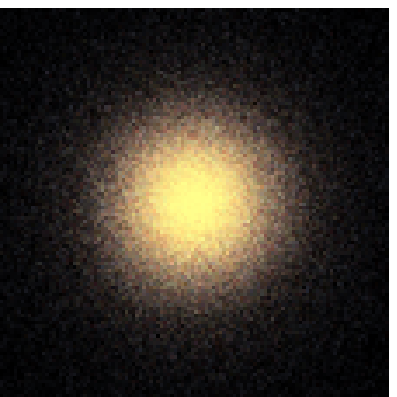,height=4cm}
\epsfig{figure=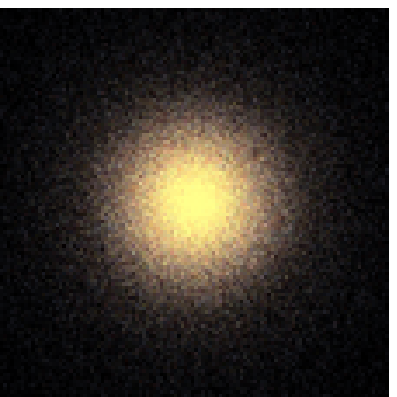,height=4cm}
\epsfig{figure=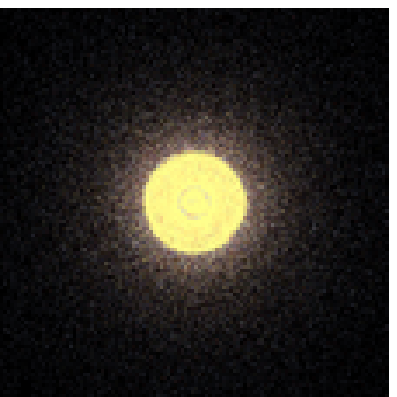,height=4cm}
\epsfig{figure=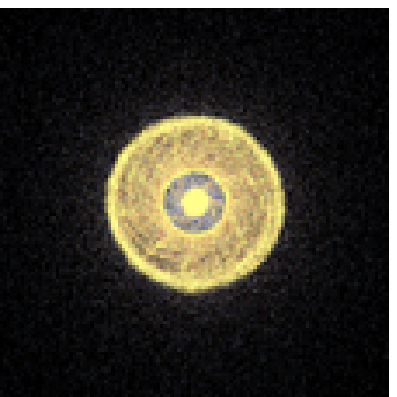,height=4cm}
\epsfig{figure=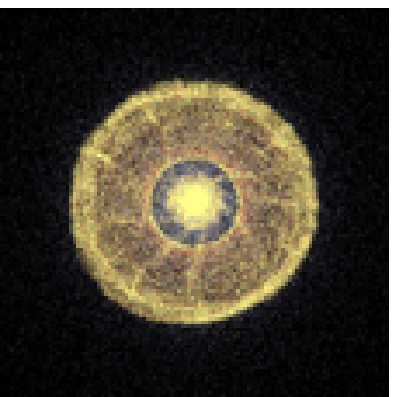,height=4cm}
\epsfig{figure=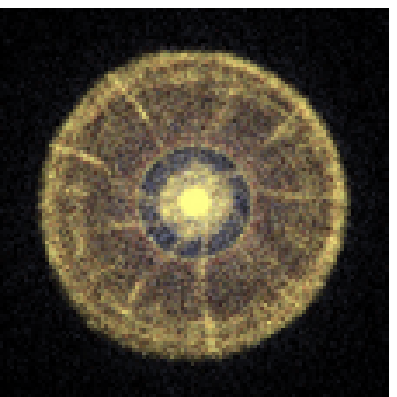,height=4cm}
}}
\caption{\label{fig:fig4} 
The same as Fig.~\ref{fig:fig3}, but Cartwheel is seen face-on.}  
\end{figure}
Figs. \ref{fig:fig1} and \ref{fig:fig2} show the evolution of run A3 (A1 and A2 are almost identical), along two different projection axes, in the first 200 Myr of the simulation. Figs. \ref{fig:fig3} and \ref{fig:fig4} show the same for run B3.
First of all, let us consider run A3. The intruder approaches Cartwheel with a non-zero inclination angle ($\sim{}14^\circ{}$). Even before the passage through the disc ($\lesssim{}40$ Myr after the beginning of the simulation), the bulge and the central part of the disc of Cartwheel are deformed by the gravitational influence of the intruder. After the encounter, a fast density wave propagates from the centre of the galaxy towards the periphery. The wave does not remain on the plane of the galaxy, but tends to be deflected towards the intruder, generating a bell-like edge-on galaxy.

After $\sim{}120$ Myr from the beginning of the simulation, the density wave forms a well-defined propagating ring. Interestingly, the ring is not perfectly circular, but slightly elongated, because of the initial inclination angle between Cartwheel and the intruder (see also Horellou \& Combes 2001). This indicates that the difference between major and minor axis in Cartwheel, normally ascribed to the inclination angle of the disc with respect to the observer ($\sim{}40^\circ{}$, Higdon 1996), can be an intrinsic deformation due to the interaction itself. 
An intrinsic asymmetry of the ring (rather than an effect of inclination) is somehow supported also by observations, which indicate that most of the H$\alpha{}$ (Higdon 1995) and X-ray emission (Gao et al. 2003; Wolter \& Trinchieri 2004) is concentrated in the southern quadrant of the ring.
The inner, less-developed ring is also well visible in the simulations.

\begin{figure}
\center{{
\epsfig{figure=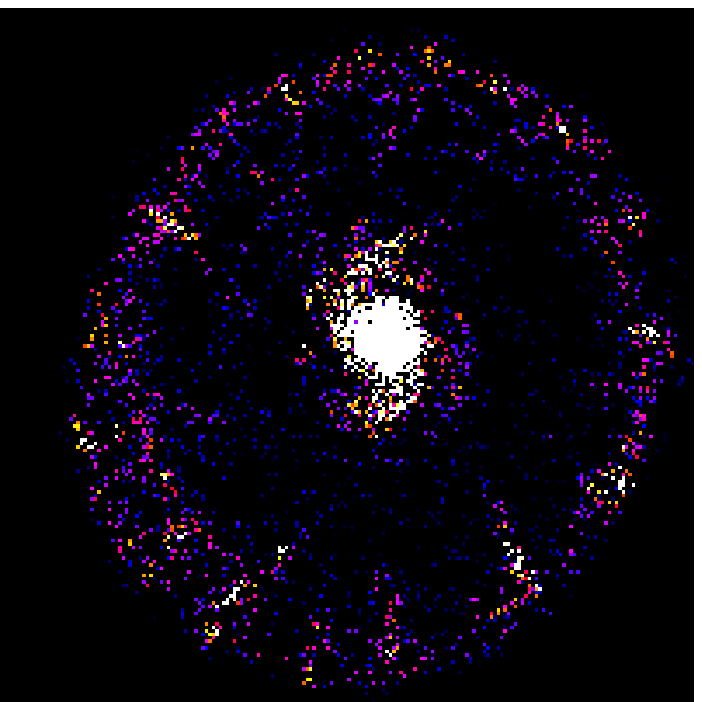,height=8cm}
\epsfig{figure=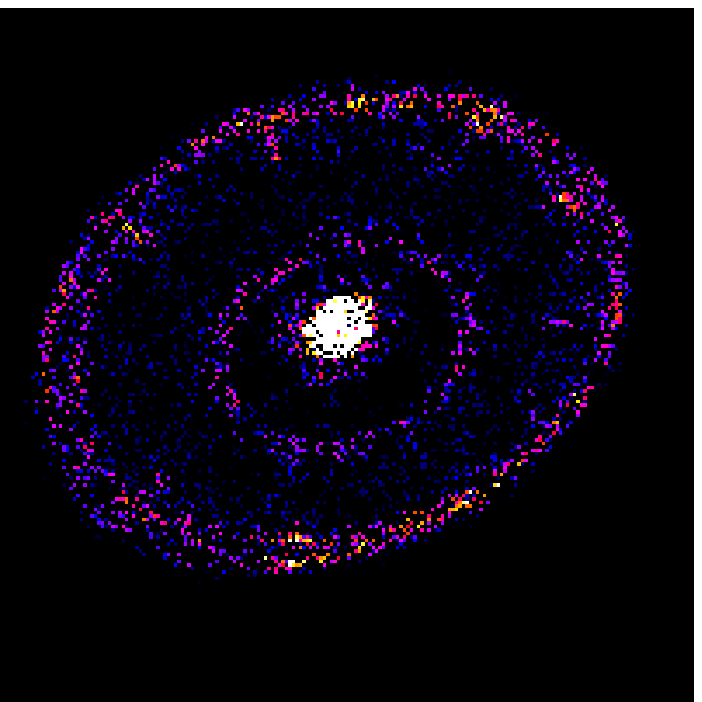,height=8cm}
}}
\caption{\label{fig:fig5}
Density of the gas in run A3 (at $t=140$ Myr, top panel) and B3 (at $t=160$ Myr, bottom panel). 
The color coding indicates the projection along the z axis of the gas density in linear scale (from 0 to 20 $M_\odot{}$ pc$^{-2}$). Each frame measures 70 kpc per edge.
} 
\end{figure}

Another interesting feature of runs A1$-$A3 is the presence of well-developed spokes after $\sim{}120$ Myr from the beginning of the simulation. The upper panel of Fig.~\ref{fig:fig5} (where only the gas component is shown, at a time  $t=$140 Myr from the beginning of run A3) and Fig.~\ref{fig:fig6}  (where only stars are plotted, at  $t=$140 Myr) show that these spokes are mainly composed by stars and are quite gas-poor. This result perfectly agrees with observations: spokes appear bright in red continuum (Higdon 1995), but hardly detectable in H$\alpha{}$ (Higdon 1995), 21 cm (Higdon 1996) and radio continuum (Mayya et al. 2005). The spokes in our simulations probably originate  from gravitational instabilities in the stellar ring (see Appendix A for more details on the formation of spokes and for a comparison with previous studies).

The time evolution of run B3 (Figs. \ref{fig:fig3} and \ref{fig:fig4}) is quite similar to run A3. However, the velocity of the intruder is slightly higher with respect to the escape velocity from Cartwheel, so that the bell-like shape of the edge-on post-encounter Cartwheel is less pronounced: the galaxy remains almost in the initial plane, even if the disc becomes thicker towards the direction of the intruder.
In this case, the intruder approaches Cartwheel along its symmetry axis. Then, the rings which propagate after the encounter are perfectly circular. To obtain the observed ratio between the two axes of Cartwheel, we have to rotate the simulated face-on galaxy of $\sim{}37^{\circ{}}-40^{\circ{}}$, as it has been done in the bottom panel of Fig.~\ref{fig:fig5} and in Fig.~\ref{fig:fig7}. 

The stellar spokes appear also in run B3, but are much less pronounced than in run A3, indicating that they depend on the strength of the interaction and especially on the total mass in stars. We made other check runs, to understand for which parameters the spokes form and how frequent is their production. Most of runs show only underdeveloped spokes, like run B3. In particular, the spokes tend to disappear if we increase the central halo mass, if we reduce the mass of the disc (stabilizing it), if we increase the relative velocity between intruder and target and if we reduce the mass of the intruder. Summarizing, runs A are the only case (among $\sim{}$10 check runs with different parameters) in which spokes are so well developed. The fact that spokes seem to form only for a small range of initial parameters explains why they are observed only in $\sim{}3$ of the known ring galaxies (Higdon 1996).

Apart from these differences, both runs A and B reproduce the main features of Cartwheel. The size of the ring agrees with Cartwheel observations especially from $t\sim{}120$ to $t\sim{}200$ Myr from the beginning of the simulation (i.e. $t\sim{}80$ to $t\sim{}160$ Myr from the gravitational encounter). This result agrees with models based on the 
radial optical and near-infrared color gradients (Vorobyov \&{} Bizyaev 2003), which predict for Cartwheel an age $<250$ Myr.
The age estimated by previous simulations ($\sim{}160$ Myr from the gravitational encounter, Hernquist \& Weil 1993) is also in agreement with our findings.
In the following, we will focus on runs A at $t=140$ Myr and runs B at  $t=160$ Myr, as  fiducial ages.

 The only feature of our simulations which is  not in perfect agreement with the observations is the presence of gas at the centre of Cartwheel, both in runs A (upper panel of Fig.~\ref{fig:fig5}) and B (bottom panel). Observations do not show significant H$\alpha{}$  (Higdon 1995) or HI emission lines (Higdon 1996) from the centre and the inner ring. In particular, the HI mass within 22 and 43 arcsecond (i.e. $\sim{}13$ and $\sim{}25$ kpc) is $\sim{}10^8$ and $\sim{}1.5\times{}10^9\,{}M_\odot{}$, respectively (Higdon 1996). These observational estimates are a factor of $\sim{}5$ smaller than the values in our simulations.

However, previous simulations (Hernquist \& Weil 1993; Mihos \& Hernquist 1994; Horellou \& Combes 2001) agree with our results, showing that a large fraction of gas must end up at the centre. Thus, what kind of process can either lead  to the removal of the central gas or stop the SF activity? Previous papers consider the intruder as either a rigid body (Horellou \& Combes 2001) or a merely dark matter object (Hernquist \& Weil 1993; Mihos \& Hernquist 1994). One can wonder whether a gas-rich companion can strip the gas from the centre of Cartwheel. However, we made a test run where the companion has an exponential disc of gas (total mass $2\times{}10^9M_\odot{}$, consistent with the observations of G3, Higdon 1996), and we did not find any substantial difference. 

Another hypothesis is that the effect of efficient SF (exhausting  gas in the centre before than in the ring) has exhausted the central gas. 
As it will be discussed in Section 4.2.2, the efficiency of SF in the centre is very high, especially in the first stages of the Cartwheel formation. However, we still have gas in the centre after $\sim{}$300 Myr. Maybe, there are deviations from the Schmidt law in such a peculiar environment (Higdon 1996; Vorobyov 2003).  Thus, a better recipe for SF, gas cooling and feedback might attenuate this problem.

 Furthermore, $Chandra$ X-ray observations (Wolter \& Trinchieri 2004) indicate the presence of a diffuse component in the centre of Cartwheel, which can be fit by a power law plus a plasma model. Even if the power law is associated with unresolved binaries, the  plasma component is likely due to hot diffuse gas and its total luminosity ($\sim{}3\times{}10^{40}$ erg s$^{-1}$) is comparable with the soft gaseous component of most X-ray luminous starburst galaxies\footnote{Recent {\it XMM-Newton} observations (A. Wolter, private communication) suggest that the total luminosity due to hot diffuse gas is slightly lower ($\sim{}2\times{}10^{40}$ erg s$^{-1}$, of which one-third is due to the centre and inner ring) than the value previously derived by $Chandra$ (Wolter \& Trinchieri 2004).}. 

 This suggests that the high temperature of the gas at the centre of Cartwheel (probably due to stellar or BH feedback) has stopped SF, while our simulations are not able to account for this physical process.

In addition, Horellou et al. (1998) detected CO emission from Cartwheel, indicating the presence of $\sim{}1.5\times{}10^9\,{}M_\odot{}$ of molecular gas in the inner $\sim{}25$ kpc of Cartwheel (but most of the detected molecular gas is probably concentrated in the inner $\sim{}13$ kpc). 
Thus, the centre of Cartwheel, although lacking of atomic cold gas (responsible of the HI emission line) and of SF activity (responsible of the H$\alpha{}$ emission), seems to host a conspicuous amount of hot gas and also a significant mass of molecular gas.

 Moreover, H$\alpha{}$ observations (Higdon \& Wallin 1997) 
show that the nucleus of AM 0644$-$741, a ring galaxy similar to Cartwheel, is quite gas-rich. Thus, the dearth of cold gas in the nucleus of Cartwheel could be a peculiarity of this galaxy, due to some particular stage of its evolution (e.g. feedback from the central BH or SF).


\subsection{Dynamics of IMBHs}

\begin{figure}
\center{{
\epsfig{figure=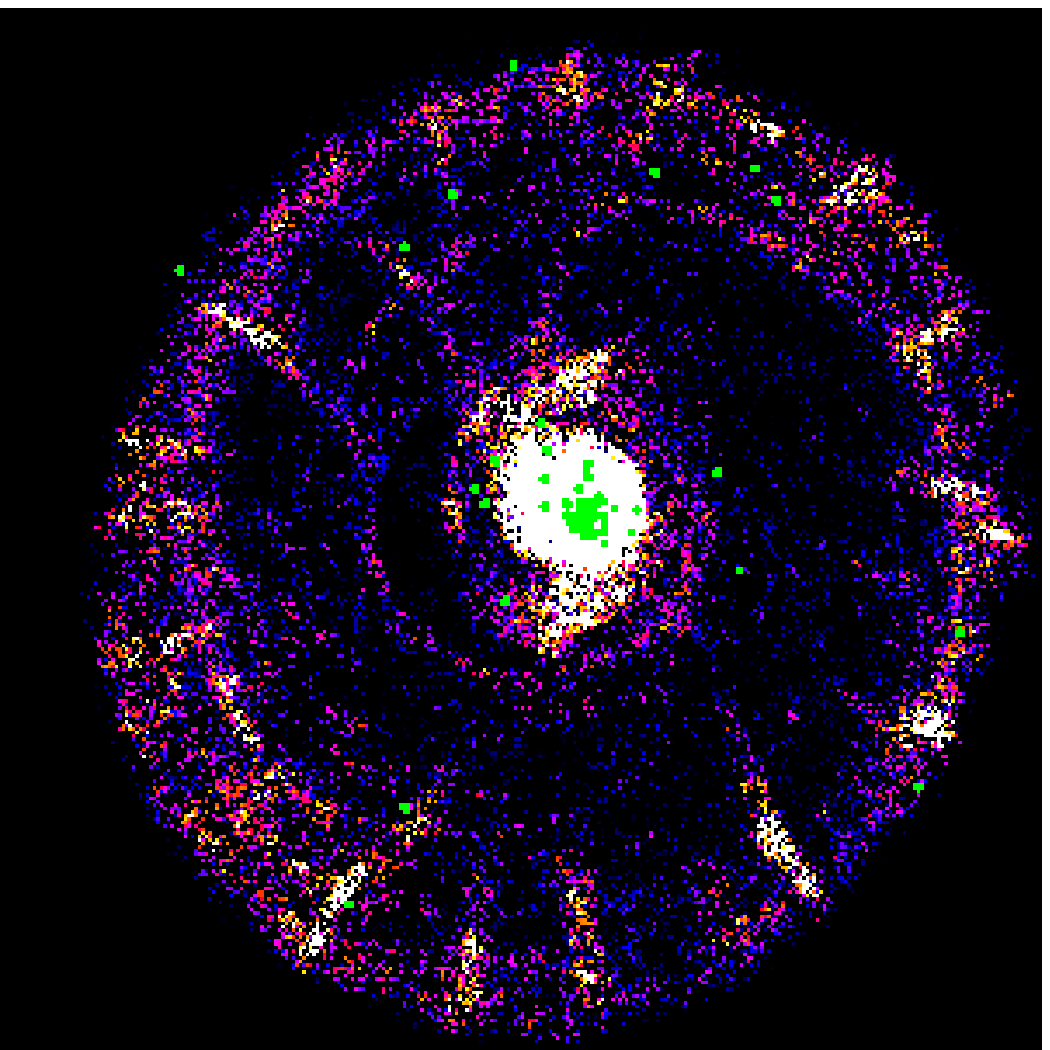,height=8cm}
\epsfig{figure=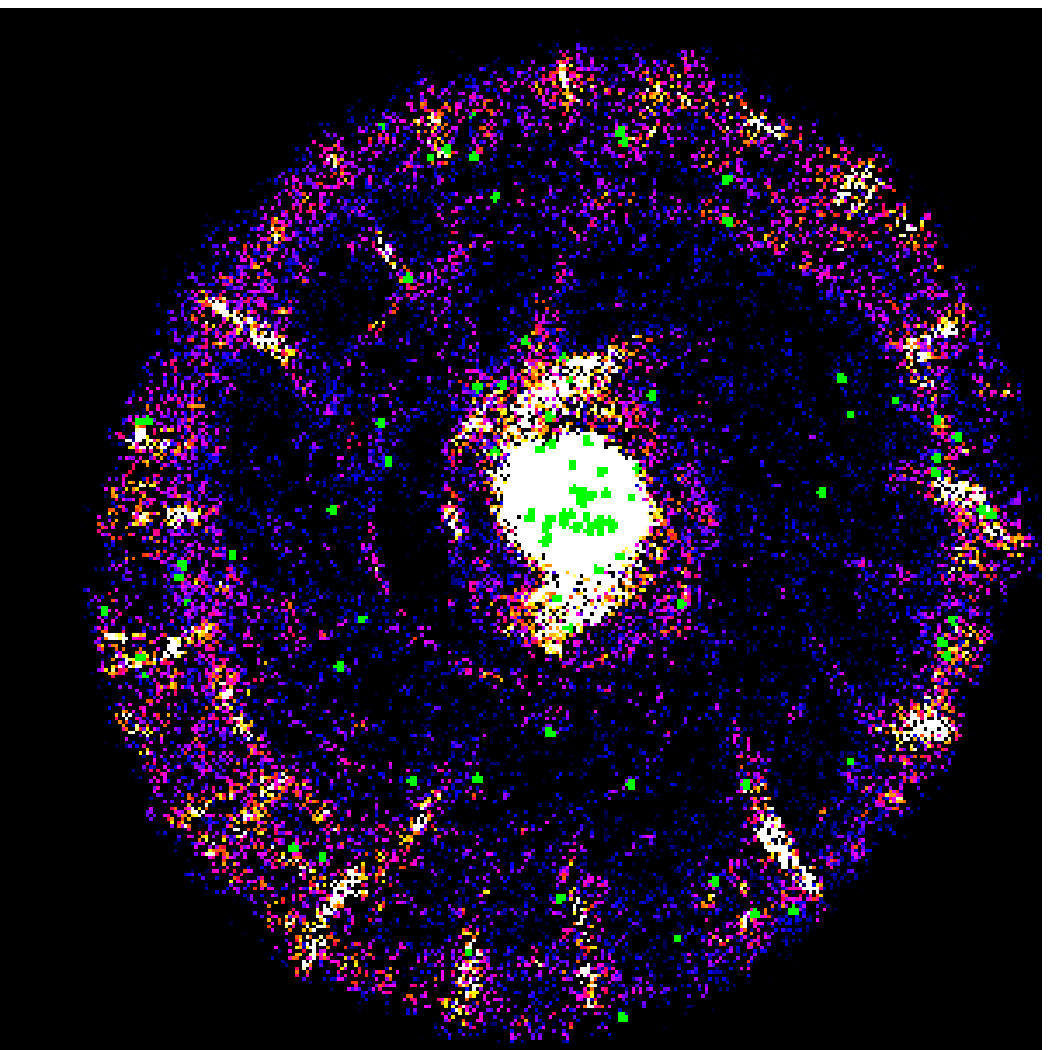,height=8cm}
}}
\caption{\label{fig:fig6} 
Density of stars at $t=140$ Myr in run A1 (top panel) and A3 (bottom panel). 
The color coding indicates the projection along the z axis of the star density in linear scale (from 0 to 40 $M_\odot{}$ pc$^{-2}$).  The filled circles (green in the online version) mark the IMBH particles. Each frame measures 70 kpc per edge.
} 
\end{figure}

\begin{figure}
\center{{
\epsfig{figure=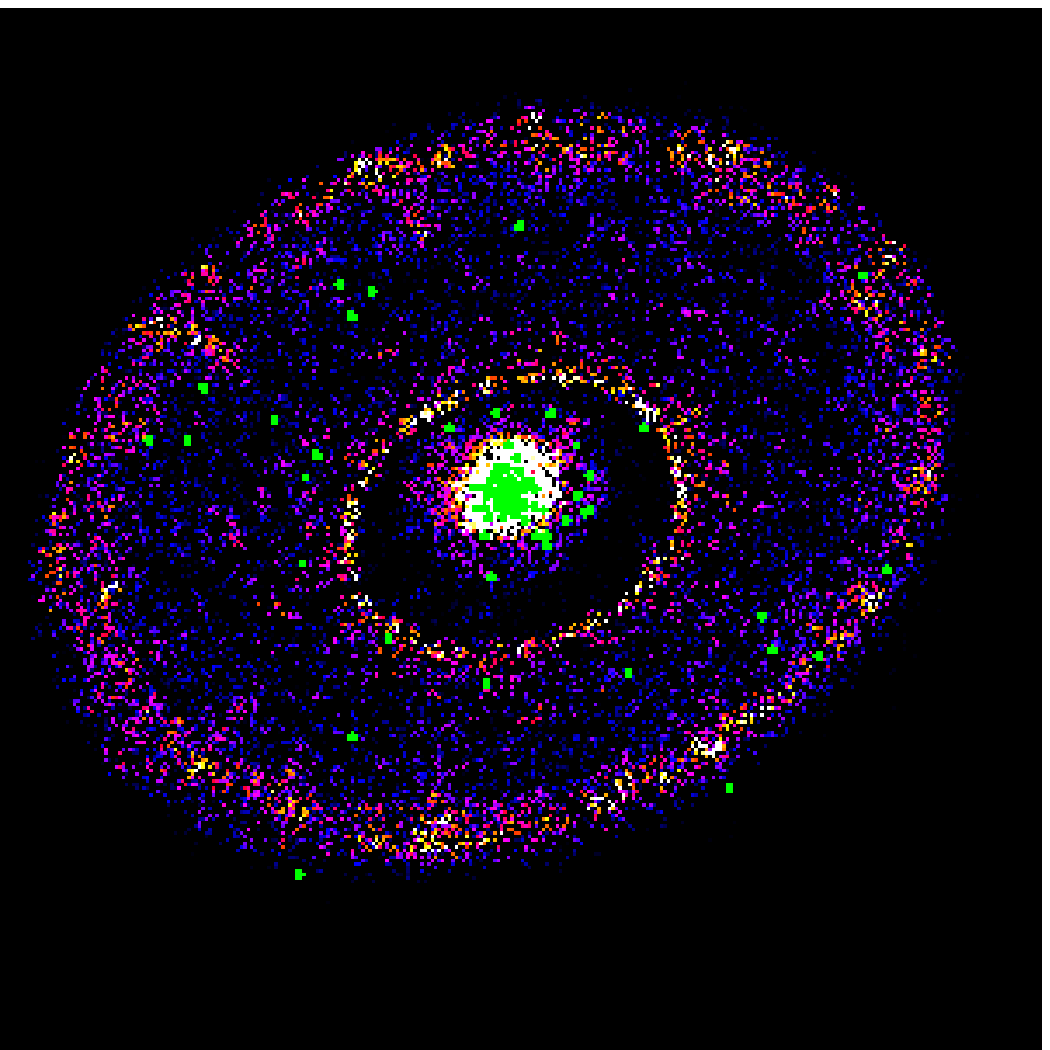,height=8cm}
\epsfig{figure=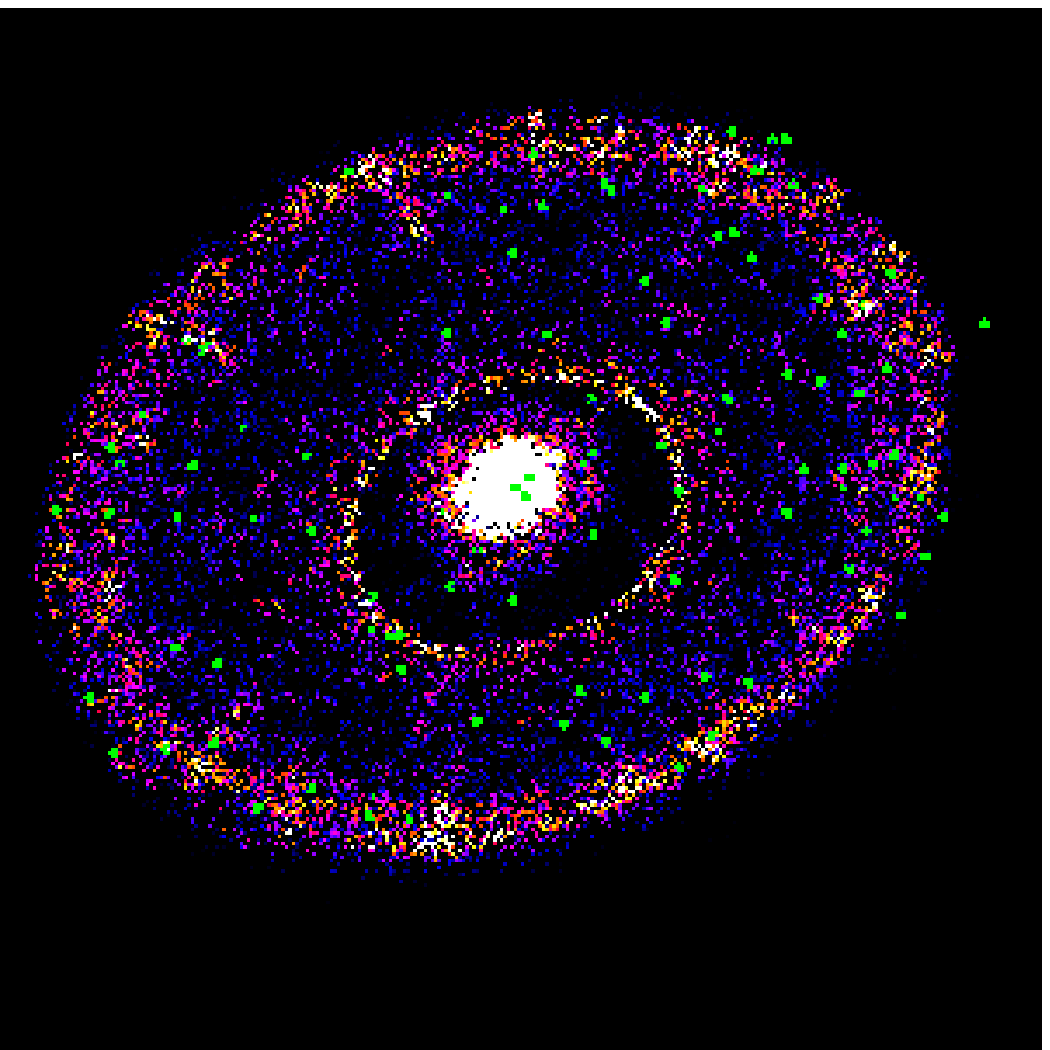,height=8cm}
}}
\caption{\label{fig:fig7} 
Density of stars at $t=160$ Myr in run B1 (top panel) and B3 (bottom panel). 
The color coding indicates the projection along the z axis of the star density in linear scale (from 0 to 20 $M_\odot{}$ pc$^{-2}$).  The filled circles (green in the online version) mark the IMBH particles. Each frame measures 70 kpc per edge.
} 
\end{figure}

Let us consider now the dynamics of the IMBHs during the formation of the ring. In Fig.~\ref{fig:fig6} we marked the positions of IMBH particles at $t=140$ Myr in the case of runs A1 (top panel) and A3 (bottom panel).  Fig.~\ref{fig:fig7} shows the same at $t=160$ Myr for runs B1 (top panel) and B3 (bottom panel). 

In both cases, halo IMBHs appear much more concentrated than disc IMBHs. As we would have expected, disc IMBHs behave like stars (having the same initial distribution), and an important fraction of them ends up inside the outer ring. Instead, halo IMBHs are only slightly perturbed by the intruder.

\begin{figure}
\center{{
\epsfig{figure=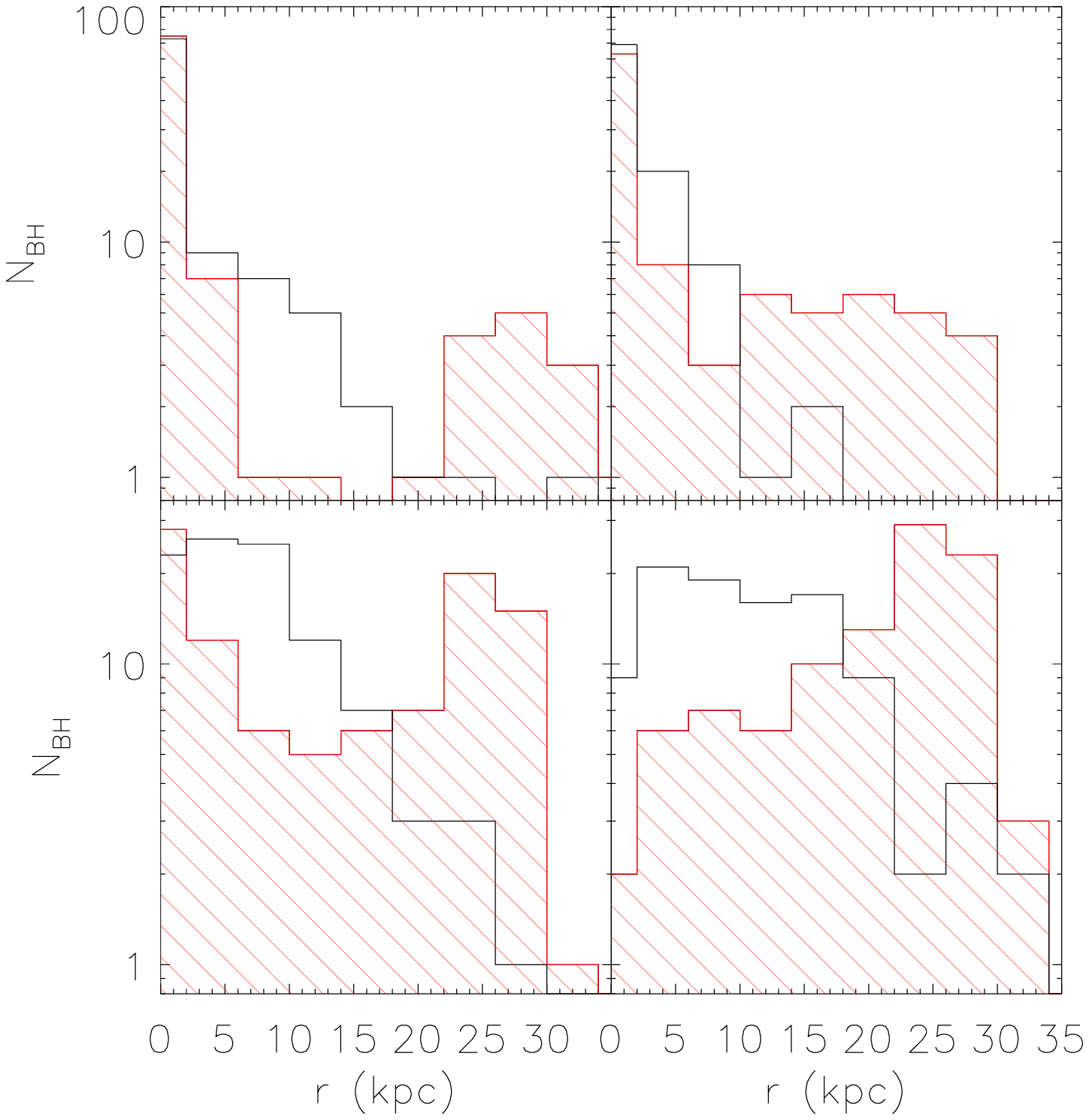,height=8cm}
}}
\caption{\label{fig:fig8} 
Distribution of IMBHs as a function of radius. Top row: run A1 (left panel) and B1 (right). Bottom row: run A3 (left) and B3 (right). Empty histograms show the initial conditions, hatched histograms show the distribution at $t=140$ Myr (runs A1 and A3) or $t=160$ Myr (runs B1 and B3).
} 
\end{figure}


This qualitative consideration can be quantified by looking at Fig.~\ref{fig:fig8}, where the radial distribution of IMBHs is shown for runs A1, A3, B1 and B3, and at the third column of Table 2, where the number of IMBHs within the ring is listed.
The distribution of IMBHs after $t=140-160$ Myr is never the same as the initial one. Even some tens of halo IMBHs (runs A1 and B1) are strongly perturbed by the galaxy interaction: their final orbits are up to 30 kpc far from the galactic centre.
However, the orbits of disc IMBHs (runs A3 and B3) are far more perturbed: more than half of the BHs are dragged into the outer ring.

\section{Are the X-ray sources associated with IMBHs?}
As we have shown in the previous Section, after the gravitational interaction a fraction of IMBHs is pulled into the outer ring. The main question we would like to address is whether these IMBHs can power all or a part of the ULXs observed in Cartwheel.
IMBHs can switch on as X-ray sources both by gas accretion and by tidal capture of stars. In this Section we will consider both mechanisms.
\subsection{IMBHs accreting gas}
The details of gas accretion by IMBHs are basically unknown. As the simplest approximation (Mii \& Totani 2005; Mapelli et al. 2006; Mapelli 2007), we can assume that these IMBHs accrete at the Bondi-Hoyle luminosity\footnote{$L_X$ in equation~(\ref{eq:bondihoyle}) indicates the total X-ray luminosity. Properly speaking, the Bondi-Hoyle formula refers to the bolometric luminosity. However, in the case of ULXs the X-ray luminosity is much higher than the optical luminosity (Winter, Mushotzky \& Reynolds 2006), justifying our approximation.}:
\begin{equation}\label{eq:bondihoyle}
L_X(\rho{}_g,\,{}v)=4\,{}\pi{}\eta{}\,{}c^2\,{}G^2\,{}m_{\rm BH}^2\,{}\rho{}_g\,{}\tilde{v}^{-3},
\end{equation}
where $\eta{}$ is the radiative efficiency, $c$ the speed of light, $G$ the gravitational constant, $m_{\rm BH}$ the IMBH mass, $\rho{}_g$ the density of the gas surrounding the IMBH. $\tilde{v}=(v^2+\sigma{}_{MC}^2+c_s^2)^{1/2}$, where $v$ is the relative velocity between the IMBH and the gas particles, $\sigma{}_{MC}$ and $c_s$ are the molecular cloud turbulent velocity and gas sound speed, respectively. 

 Due to resolution limits\footnote{It is worth reminding that the Bondi-Hoyle formula refers only to gas well within the influence radius of the BH. Furthermore, the efficiency of the accretion is strongly dependent on the angular momentum distribution of the gas inside this influence radius (Agol \& Kamionkowski 2002). We do not have sufficient resolution to account for this physics. Thus, we are interested only to derive an upper limit of $L_X$.}, we cannot account for the local properties of the gas. In particular, the local density of the gas never reaches the values expected for molecular clouds. But we know from the observations (Higdon 1995) that Cartwheel ring  should be rich of molecular clouds. Thus, if we calculate $\rho{}_g$ directly from our simulation, we probably underestimate the accretion rate.

Then, as an upper limit, we assume that an IMBH passing through the ring 'intercepts' a molecular cloud whenever its distance from the closest gas particle is less than  $r_g=0.5$ kpc (i.e. 50 times the softening length). If this occurs, we calculate $L_X$ by assuming $\rho{}_g=n_{mol}\,{}\mu{}\,{}m_{\rm H}$, where $\mu{}\sim{}2$ is the molecular weight, $m_{\rm H}=1.67\times{}10^{-24}$ g is the proton mass and $n_{mol}=10^2$ cm$^{-3}$ is the mean density of a molecular cloud (Sanders, Scoville \& Solomon 1985).

$v$ and $c_s$ are extracted directly from our simulations, adopting the same technique used in Mapelli (2007). 

In particular,  $v$ is derived as the average  relative velocity between the IMBH and the gas particles which are within a distance $r_g$ from the IMBH.
Similarly, the sound speed of the gas, $c_s$, around each IMBH is calculated as the average sound speed of gas particles within the same radius $r_g$, by using the relation $c_s^2=2\,{}u_g$ (where $u_g$ is the average internal energy per unit mass of the gas particles within $r_g$). Finally,  we adopt the average observed Galactic value $\sigma{}_{MC}=3.7$ km s$^{-1}$ (Mapelli et al. 2006 and references therein).
\begin{figure}
\center{{
\epsfig{figure=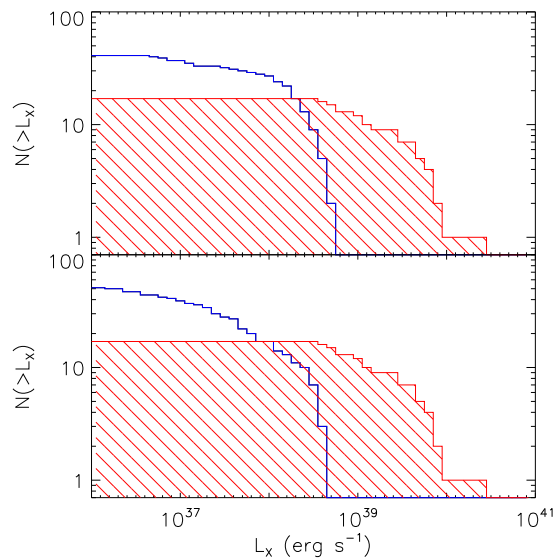,height=8cm}
}}
\caption{\label{fig:fig9} 
Cumulative X-ray luminosity observed in Cartwheel (shaded histograms) compared with simulations (open histograms). The simulated sources are obtained assuming IMBHs of mass $m_{\rm BH}=10^3\,{}M_\odot{}$, accreting gas with efficiency $\eta{}=0.1$.  Upper panel: run A3. Lower panel: run B3.  
} 
\end{figure}

Fig.~\ref{fig:fig9} shows the X-ray luminosity distribution of disc IMBHs for runs A3 (top panel) and B3 (bottom), considering only the IMBHs which are in the ring. These luminosities are obtained by applying the procedure described above and by assuming $\eta{}=0.1$ and $m_{\rm BH}=10^3\,{}M_\odot{}$. 
The simulated sources exceed the number of the observed ones at  low luminosities; but this is not significant, as observations are complete only for $L_X\gtrsim{}5\times{}10^{38}$ erg s$^{-1}$.
Instead, both in runs A3 and B3 the luminosity of IMBHs remains well below (a factor of $\gtrsim{}10$) the high-luminosity tail of the observed distribution, even with our optimistic assumptions.

Of course, if we increase  $m_{\rm BH}$ and/or $\eta{}$, $L_X$ increases according to equation (\ref{eq:bondihoyle}).
However,  $\eta{}=0.1$ and  $m_{\rm BH}=10^3\,{}M_\odot{}$ are upper limits for our model, as it is very difficult to produce IMBHs with mass larger than $\sim{}10^3\,{}M_\odot{}$  in the runaway collapse scenario (Portegies Zwart \& McMillan 2002) and it is unlikely to have higher radiative efficiency. Indeed, adopting $\eta{}=0.1$ means that we are assuming a thin disc accretion (Shakura \& Sunyaev 1973). The gas surrounding a $10^3\,{}M_\odot{}$ IMBH is unlikely to form a thin disc (Agol \& Kamionkowski 2002), as its expected accretion rate is  (Beskin \& Karpov 2005; Mapelli 2007\footnote{In Mapelli (2007) this formula is affected by a typographical error: the exponent of the velocity should be $-3$, as in our equation~(7).}):
\begin{equation}
\frac{\dot{M}}{\dot{M}_{\rm Edd}}\sim{}10^{-2}\,{}\left(\frac{m_{\rm BH}}{10^3\,{}M_\odot{}}\right)\,{}\left(\frac{n_{mol}}{10^2\,{}{\rm cm}^{-3}}\right)\,{}\left(\frac{\tilde{v}}{40\,{}{\rm km}\,{}{\rm s}^{-1}}\right)^{-3},
\end{equation}
where $\dot{M}_{\rm Edd}$ is the Eddington accretion rate.
For such low accretion rates an advection-dominated accretion flow is likely to establish, with radiative efficiency a factor of $\sim{}100$ lower than in the thin disc model (Narayan, Mahadevan \& Quataert 1998). Thus, assuming $\eta{}=0.1$ is a generous upper limit.

We do not even show the plot of halo IMBHs, i.e. runs A1 and B1, because they present only 1 and 2 X-ray sources, respectively, with luminosity lower than $10^{37}$ erg s$^{-1}$. This is mainly due to the fact that only a small fraction of halo IMBHs passes through the disc (and close enough to gas particles) for a sufficiently long lapse of time. Furthermore, halo IMBHs, even when they pass through the disc, have high relative velocities ($v\gtrsim{}100$ km s$^{-1}$) with respect to  the closest gas particles.

Then, we can conclude that ULXs in Cartwheel can hardly be originated by gas accreting IMBHs. Indeed, our model oversimplifies the process of accretion. A better treatment of gas cooling and feedback might give different estimates. However, most of our assumptions  are generous upper limits, strengthening our result.


\subsection{IMBHs accreting from stars}
IMBHs can also accrete by mass transfer in binaries. It is unlikely that halo IMBHs can capture a star and form a binary system. Recently, Kuranov et al. (2007) found that the expected number of ULXs powered by halo IMBHs which have tidally captured a stellar companion is  $\sim{}10^{-7}$ per galaxy (or $0.01$ sources per galaxy with $L_X\gtrsim{}10^{36}$ erg s$^{-1}$). Then, in this Section, we will not consider halo IMBHs.

Instead, IMBHs born by runaway collapse (i.e. disc IMBHs) are hosted in star clusters, and have a significant probability of being in binary systems.  It is also likely that IMBHs remain inside the parent cluster during its lifetime (Colpi, Mapelli \& Possenti 2003).
However,  some observed ULXs have been found displaced from the star clusters (e.g. in the Antennae, see Zezas et al. 2002),  suggesting that the BHs powering these ULXs and their companion stars have been ejected from the parent cluster.
In this paper, we assume that all the IMBHs remain in their parent cluster, which gives us an upper limit for the mass-transfer time [see equation (\ref{eq:oldstars})].



 X-ray sources originated by accreting  $\sim{}100-1000\,{}M_\odot{}$ IMBHs are expected to be transient, if the mass of the companion is $\lesssim{}10\,{}M_\odot{}$ (Portegies Zwart et al. 2004). The duty cycle shows a short (few days) bright phase followed by a rather long (several weeks) quiescent phase. The peak luminosity of these transient sources is always in the ULX range ($L_X>10^{40}$ erg s$^{-1}$, Portegies Zwart et al. 2004).

Instead, Patruno et al. (2005) showed that  $\sim{}1000\,{}M_\odot{}$ IMBHs accreting from high-mass ($\gtrsim{}10-15\,{}M_\odot{}$) stellar companions should produce persistent X-ray sources, with luminosities ranging from $10^{36}$ erg s$^{-1}$ to more than $10^{40}$ erg s$^{-1}$.

Of the 17 point X-ray sources (with $L_X>10^{38}$ erg s$^{-1}$) associated with Cartwheel (see Fig.~\ref{fig:fig9} and Wolter \& Trinchieri 2004),  at least 3 are variable over a time-scale of 6 months (Wolter et al. 2006). Other 4 sources could be constant, at least  over a time-scale of 6 months, as they have not shown variability in the $XMM-Newton$ observations reported by Wolter et al. (2006). Also the variable sources are not properly transient (e.g. the source N.10 dims of a factor of $\sim{}2$ over a time-scale of 6 months, after having been constant for at least 4 years). Thus, in Cartwheel both variable and constant sources can exist; but there are not evidences for transient sources (where for transient we mean sources with a bright phase of a few days and a long quiescent phase).

 We also point out that it is physically unlikely that all of these 17 point sources are powered by IMBHs. Among them only the source N.10 has a luminosity higher than $10^{41}$ erg s$^{-1}$ (Gao et al. 2003; Wolter \& Trinchieri 2004), which can hardly be explained with a BH mass smaller than $100\,{}M_\odot{}$.
Other 4 sources have $L_X\gtrsim{}5\times{}10^{39}$ erg s$^{-1}$, which is very high but still consistent with models of X-ray sources powered by BHs with mass $\lesssim{}100\,{}M_\odot{}$. The remaining sources have  $L_X\lesssim{}5\times{}10^{39}$ erg s$^{-1}$. Thus, apart from the source N.10, all the X-ray sources in Cartwheel are perfectly consistent with various models of ULXs powered by stellar mass  or moderately massive ($\lesssim{}100\,{}M_\odot{}$) BHs (e.g. via super-Eddington accretion, King \& Pounds 2003).

In this Section we want to address whether it is theoretically possible that all these ULXs or at least the brightest among them are powered by IMBHs accreting by mass transfer from stellar companions.

In order to check this, we calculate the number of X-ray sources which are powered by our simulated IMBHs. Firstly, we consider the case in which IMBHs accrete only from old stars, then we focus on IMBHs accreting from young stars.
\begin{table}
\begin{center}
\caption{IMBHs in the ring.
}
\begin{tabular}{llll}
\hline
\vspace{0.1cm}
Run &  time (Myr) & $N_{\rm BH,\,{}ring}$$^a$ &  $N_{\rm BH,\,{}SF}$$^b$\\
\hline
A1  &  140 & 17 & -\\ 
A2  &  140 & 50 & -\\ 
\vspace{0.1cm}
A3  &  140 & 50 & 27\\ 
B1  &  160 & 21 & -\\ 
B2  &  160 & 79 & - \\ 
B3  &  160 & 79 & 30 \\ 
\hline
\end{tabular}
\end{center}
{\footnotesize $^{a}$Number of IMBHs within (or close to) the ring, i.e. more than 15 kpc away from the centre of the galaxy.\\
$^{b}$Number of IMBHs within (or close to) the ring which are close to a star-forming region.}\\
\label{tab_2}
\end{table}

\subsubsection{IMBHs accreting from old stars}
Blecha et al. (2006) found that a $100-500\,{}M_\odot{}$ IMBH located in a young cluster spends $\sim{}3$ per cent of its life in the mass-transfer phase with a stellar companion. So, the number ($N_{\rm BH,\,{}MT}$) of $\sim{}100\,{}M_\odot{}$ IMBHs, born before the dynamical interaction,   which are accreting by mass transfer from old stars, in the ring, at the current time, will be simply given by
\begin{equation}\label{eq:oldstars}
N_{\rm BH,\,{}MT}= 2.4\,{}\left(\frac{f_{\rm MT}}{0.03}\right)\,{}\left(\frac{N_{\rm BH,\,{}ring}}{79}\right),
\end{equation}
where $f_{\rm MT}$ is the fraction of the star cluster life in which the IMBH undergoes mass transfer (from Blecha et al. 2006), and $N_{\rm BH,\,{}ring}$ is the number of IMBHs in the stellar ring (see column 3 of Table~2). This calculation implies that in runs A2$-$A3 (B2$-$B3) there are only 1.5 (2.4) IMBHs undergoing mass transfer. 

Furthermore, if both the IMBH and the stellar companion formed before the dynamical interaction between Cartwheel and the intruder, we expect that the companion is older than $\sim{}100$ Myr, which is approximately the lifetime of 6-7 $M_\odot{}$ stars. Then, X-ray sources due to IMBHs and stars born before the dynamical interaction of Cartwheel are expected to be transient. 
In this case, the time spent in the outburst phase is probably only a few per cent of the total mass-transfer time (McClintock \&{} Remillard 2006), further reducing the probability of observing such sources (Blecha et al. 2006).

 Therefore, we have to hypothesize that the non-transient X-ray sources in Cartwheel, if their are powered by IMBHs, are due either to pre-interaction or to post-interaction born IMBHs accreting from young ($<100$ Myr) and massive ($>10\,{}M_\odot{}$) stars. 

\subsubsection{IMBHs accreting from young stars}
We checked the hypothesis of IMBHs accreting from young ($<100$ Myr) stars by setting up  SF in two of our runs (A3 and B3). The recipe for SF in our code is simply based on the Schmidt law, and we assume a SF efficiency $c_\ast{}=0.1$. This assumption results in a SF rate (in the ring) of $\sim{}36\,{}M_\odot{}$ yr$^{-1}$, in good agreement with observations (Marston \& Appleton 1995; Mayya et al. 2005).
\begin{figure}
\center{{
\epsfig{figure=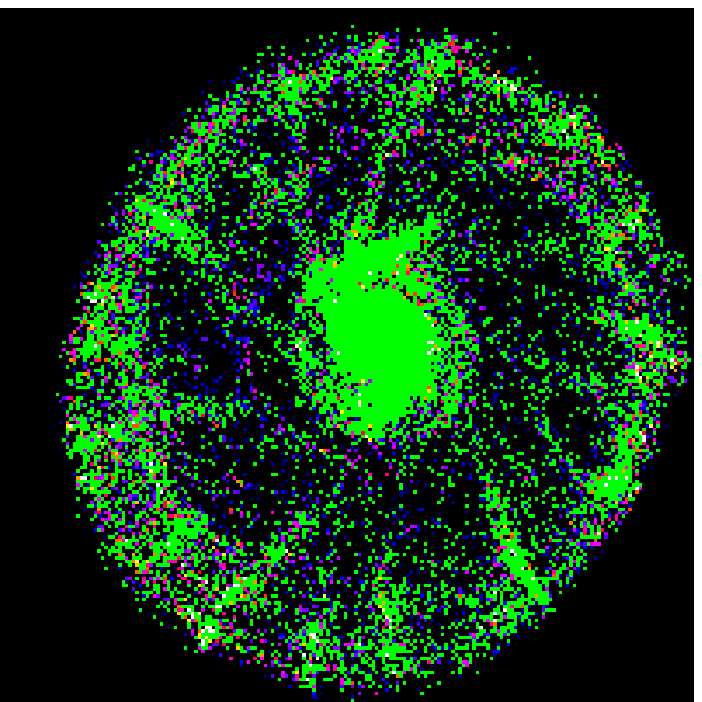,height=8cm}
\epsfig{figure=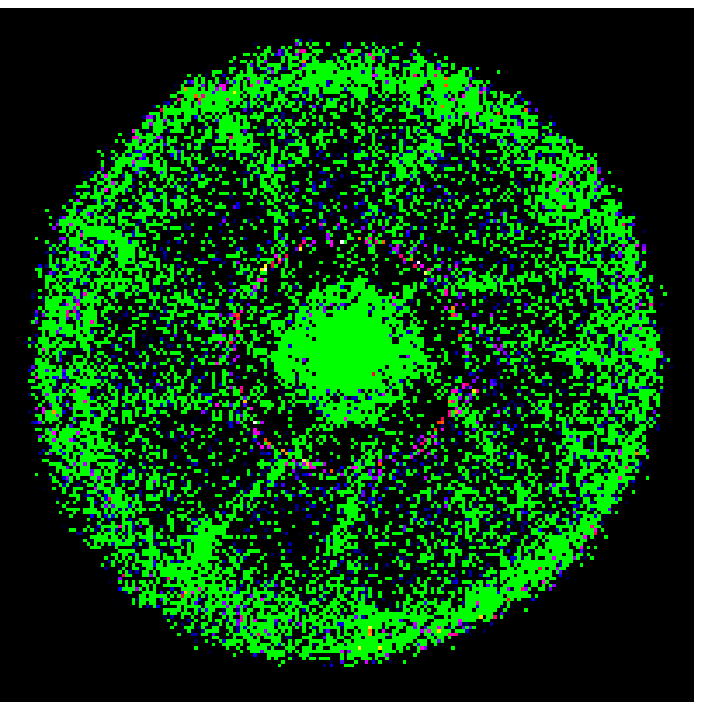,height=8cm}
}}
\caption{\label{fig:fig10} 
Newly formed stars (filled circles, green in the online version) are superimposed to the density map of old stars (the same as in Figs.~\ref{fig:fig6} and \ref{fig:fig7}). Top panel: run A3 at $t=140$ Myr. Bottom panel: run B3 at $t=160$ Myr. Each frame measures 70 kpc per edge.
} 
\end{figure}
Fig.~\ref{fig:fig10} shows stars which formed after the beginning of simulations in runs A3 ($t=140$ Myr, top panel) and B3 ($t=160$ Myr, bottom).
In both cases SF is strong in the outer ring and at the centre. While SF in the ring is consistent with observations, SF in the centre is puzzling, as no H$\alpha{}$ emission is detectable now from the centre. This problem of inconsistency between data and simulations has already been pointed out by Higdon (1996), who proposed a possible deviation in Cartwheel from the Schmidt law. On the other hand, Fig.~\ref{fig:fig11}, where successive epochs of SF in Cartwheel are shown, indicates that SF is particularly intense in the central region immediately after the interaction with the intruder (top panels), when the density wave passes through the inner 10-20 kpc. After that phase ($t>120$ Myr, bottom panels), SF proceeds especially in the propagating ring and at the very centre, while it is quenched in the intermediate parts (the region of the spokes and the inner ring). 

\begin{figure}
\center{{
\epsfig{figure=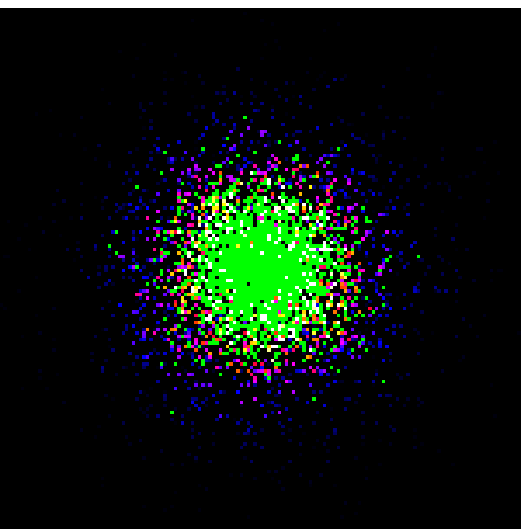,height=4cm}
\epsfig{figure=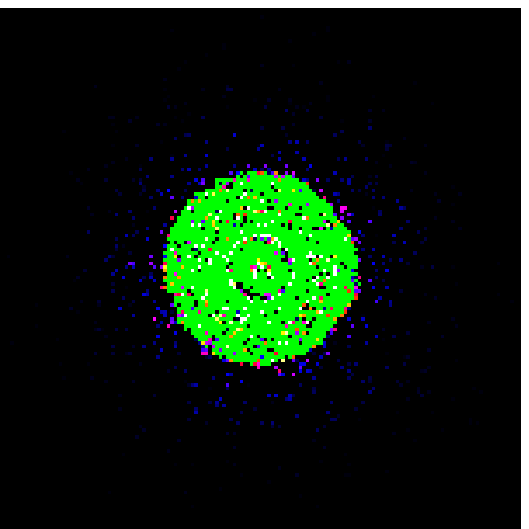,height=4cm}
\epsfig{figure=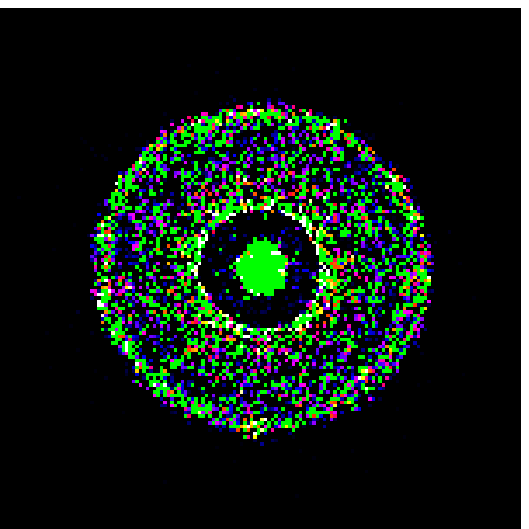,height=4cm}
\epsfig{figure=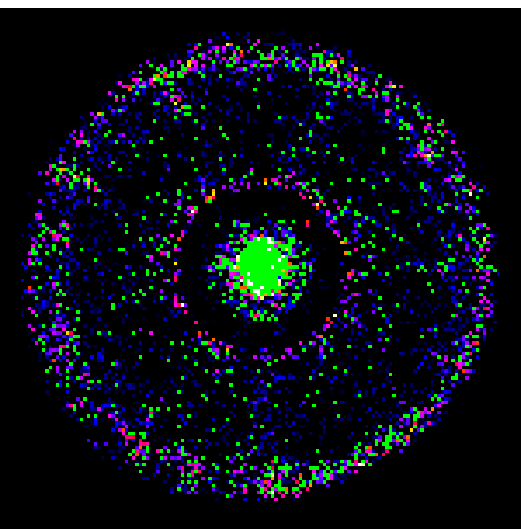,height=4cm}
}}
\caption{\label{fig:fig11} 
Newly formed stars (filled circles, green in the online version) are superimposed to the density map of old stars (the scale is the same as Fig.~\ref{fig:fig7}) for run B3. Top left panel: situation at $t=40$ Myr; filled circles are stars born at $t\leq{}40$ Myr.  Top right panel: situation at $t=80$ Myr; filled circles are stars born at $40\leq{}t\leq{}80$ Myr.  Bottom left panel: situation at $t=120$ Myr;  filled circles are stars born at $80\leq{}t\leq{}120$ Myr.  Bottom right panel: situation at $t=160$ Myr;  filled circles are stars born at $120\leq{}t\leq{}160$ Myr.
Each frame measures 70 kpc per edge.
} 
\end{figure}
Another important feature of Fig.~\ref{fig:fig10} (especially the bottom panel, run B3) is that relatively young stars appear concentrated along the spokes. As this concentration of young stars within the spokes is not evident in the frames of Fig.~\ref{fig:fig11}, we infer that most of the relatively young stars concentrated in the spokes are born in the central part of Cartwheel before or in the first stages of the interaction ($t\lesssim{}80$ Myr from the beginning of the simulation), and have been successively ejected in the spokes (see the Appendix). They are not born in the spokes, confirming that spokes are not star-forming regions.

Let us now estimate the importance of these newly formed stars for the accretion of IMBHs. Two scenarios (maybe coexistent) are possible: (i) the IMBHs were created before the interaction and then entered into  young star-forming regions, where they captured a stellar companion; (ii) the IMBHs have been formed in very young stellar clusters born after the dynamical interaction of Cartwheel and they accrete from stars belonging to the same parent cluster.

If the IMBHs were born before the interaction of Cartwheel with the intruder, we have to calculate how many of them happen to be in the outer ring and close enough to a young star-forming region. The fourth column of Table~2 shows how many IMBHs ($N_{\rm BH,\,{}SF}$) are in the outer ring (at time $t=140$ and $t=160$ Myr for run A3 and B3, respectively) and host at least 1 newly formed stellar particle in their neighborhood (the radius of the neighborhood is $r_g=0.5$ kpc, the same as in Section 4.1).
We assume that all these IMBHs plunged into a young star cluster and captured a young stellar companion (which is obviously an upper limit). In analogy with equation (\ref{eq:oldstars}), the number of IMBHs which are accreting by mass transfer at the current time is:
\begin{equation}\label{eq:newstars}
N_{\rm BH,\,{}MT}= 0.9\,{}\left(\frac{f_{\rm MT}}{0.03}\right)\,{}\left(\frac{N_{\rm BH,\,{}SF}}{30}\right),
\end{equation}
which means that both in run A3 and B3 only $\lesssim{}1$ ULX can form via this mechanism.

However, IMBHs can also form by runaway collapse in young clusters born after the dynamical interaction.
 Then, we can assume that a fraction of the newly born star particles hosts an IMBH. In this scenario, it is required that $\sim{}$1000 disc IMBHs are born in the young star clusters, in order to have $\sim{}30$ IMBHs undergoing mass transfer at present. If each newly formed young cluster hosts an IMBH (a quite optimistic assumption), this means that $\sim{}5-8$ per cent of the total mass of  stars formed in the ring (which is about half of the total mass of new stars, in our runs A3 and B3) is in young clusters.

If the age of the young cluster is $\lesssim{}40$ Myr, then it hosts also stars more massive than 10 $M_\odot{}$. IMBHs undergoing mass transfer in these very young star clusters can generate both transient (if the mass of the companion is $\lesssim{}10\,{}M_\odot{}$) and persistent X-ray sources (otherwise). A good fraction of these sources are expected to be in the ULX range.

Thus, if $\sim{}30$ IMBHs are undergoing mass transfer at present in the Cartwheel ring, it is natural to expect that they produce a number of bright X-ray sources comparable with observations ($\sim{}17$ X-ray sources brighter than $L_X>10^{38}$ erg s$^{-1}$; Wolter \& Trinchieri 2004).

But is it realistic to think that $\sim{}1000$ IMBHs have been formed (or pulled) in young clusters along the Cartwheel ring? 
The Milky Way probably hosts $\sim{}$100 young ($<10^7$ Myr) massive ($>10^4\,{}M_\odot{}$) star clusters\footnote{The currently detected young massive clusters are of the order of 10. However, assuming an homogeneous distribution and accounting for the fact that most of clusters are hidden by the obscuring material in the Galactic plane, the total number of young clusters in the Milky Way might be as high as 100 (Gvaramadze, Gualandris \& Portegies Zwart 2007).} (Gvaramadze, Gualandris \& Portegies Zwart 2007). Galaxies which are experiencing an epoch of intense SF  might have a larger population of young massive clusters. As the SF rate in Cartwheel at present is a factor $\sim{}80$ higher than in the Milky Way, it might be possible (although optimistic)  that Cartwheel hosts a factor of 10 more young massive clusters than the Milky Way.

If we accept the theory of runaway collapse (Portegies Zwart \& McMillan 2002), it is also possible that most young massive clusters host an IMBH at their centre. In fact, the only requirement to start runaway collapse is that the half-mass radius is from 0.3 to 0.8 pc (Gvaramadze et al. 2007), a condition which is satisfied by most Galactic young massive clusters. Thus, a number of $\sim{}1000$ young clusters (formed preferentially after the dynamical interaction with the intruder galaxy), hosting a $\sim{}100\,{}M_\odot{}$ IMBH at their centre, might produce a number of ULXs comparable with the observations. However, this result dramatically depends on the IMBH formation process that we assume, as only the runaway collapse model guarantees the formation of such a large number of disc IMBHs. 

Thus, we conclude that the hypothesis that all the X-ray sources  observed in Cartwheel are powered by $\sim{}100\,{}M_\odot{}$ disc IMBHs accreting from stars can be hardly  justified by our simulations, as it requires (i) that a huge number ($\sim{}500-1000$) of IMBHs and of young clusters  form in Cartwheel, (ii) that each young cluster hosts an IMBH (possible only, under extreme assumptions, in the runaway collapse scenario), (iii) that all the IMBHs remain in the parent cluster.

 On the other hand, the observational properties of ULXs suggest that only the brightest among them require the presence of an IMBH. In particular, the source N.10 in Cartwheel (see Section 4.2) has $L_X\gtrsim{}10^{41}$ erg s$^{-1}$, difficult to explain with a BH mass smaller than $100\,{}M_\odot{}$. Then, we can hypothesize that only the long-term variable source N.10 is powered by an IMBH. In order to observe only one very bright  ULX, $\sim{}30$ IMBHs 
are required to form in Cartwheel ring.

Other sources among the brightest ULXs in Cartwheel might be powered by IMBHs with mass $\sim{}100\,{}M_\odot{}$. For example, other 4 point sources in Cartwheel  have  $L_X\gtrsim{}5\times{}10^{39}$ erg s$^{-1}$.
 As all the 5 brightest sources are either persistent or variable but not transient\footnote{Here for transient sources we only mean sources with bright phase of a few days and very long quiescent phase, as described in Portegies Zwart et al. (2004).} (Wolter et al. 2006),  they cannot be powered by IMBHs with low-mass stellar companions. Thus, $\sim{}100-200$ disc IMBHs, mostly born in the last $\sim{}40$ Myr, are required to explain the $\lesssim{}5$ brightest sources.


\section{Summary}

In this paper we investigated the possible connection among IMBHs and the $\sim{}17$ (Gao et al. 2003; Wolter \& Trinchieri 2004) bright X-ray sources detected in the outer ring of Cartwheel.  Recent observations show that models based on beamed emission or super-Eddington accretion in HMXBs including stellar mass BHs can explain most of ULXs, apart from the brightest ones (Roberts 2007 and references therein). However, the observations cannot definitely exclude that all the ULXs are powered by IMBHs. Thus, in our paper we checked whether the IMBHs can account for all or only for a part of the ULXs observed in Cartwheel.

We simulated the formation of a Cartwheel-like ring galaxy via dynamical interaction with an intruder galaxy. In this simulation we also integrated the evolution of 100 IMBHs particles.

We considered two different models of IMBH formation, i.e. IMBHs born as relics of population III stars (and distributed as a concentrated halo population) and IMBHs formed via runaway collapse of stars (and distributed as an exponential disc). For these models, we investigated both gas accretion from surrounding molecular clouds and mass transfer from a stellar companion. The main results of this study are the following.
\subsection{Halo IMBHs}
IMBHs born as the relics of population III stars, if they are distributed as a halo population, cannot contribute to the X-ray sources, neither via gas accretion  nor via mass transfer in binaries. In particular, the luminosity produced by halo IMBHs accreting gas is always many orders of magnitude smaller than that of the observed sources, even if we assume that halo IMBHs have a large mass ($m_{\rm BH}=10^3\,{}M_\odot{}$) and a high radiative efficiency ($\eta{}=0.1$). This is due to the fact that only a small fraction of halo IMBHs passes through the disc (only for a short lapse of time), and even these IMBHs have a high ($v\gtrsim{}100$ km s$^{-1}$) relative velocity with respect to gas particles.
Similarly, halo IMBHs cannot accrete mass from stars, as the probability that they acquire a companion is very low.
\subsection{Disc IMBHs}
 IMBHs born from the runaway collapse of stars should be a disc population. Under overoptimistic assumptions ($n_g=10^2\,{}{\rm cm}^{-3}$, $m_{\rm BH}=10^3\,{}M_\odot{}$ and $\eta{}=0.1$) these IMBHs can produce, via gas accretion, X-ray sources with $L_X\lesssim{}10^{39}$ erg s$^{-1}$,  a factor of $\sim{}10$ fainter than the brightest ULXs observed in Cartwheel. Thus, also disc IMBHs accreting gas cannot explain the observed X-ray sources in Cartwheel. Our model of gas accreting IMBHs contains many rough assumptions; but most of them are upper limits, strengthening our result. However, a more realistic treatment of the local properties of the gas would be helpful, to understand the physical mechanisms of gas accretion onto IMBHs.

On the other hand, runaway collapse born IMBHs are hosted in dense young clusters. In such environment, it is easy for the IMBH to capture a stellar companion. Blecha et al. (2006) estimated that a $\sim{}100\,{}M_\odot{}$ IMBH undergoes mass transfer from a companion star for about the $3$ per cent of the cluster lifetime. Previous papers (Portegies Zwart et al. 2004; Patruno et al. 2005) have shown that IMBHs accreting from low-mass ($\lesssim{}10\,{}M_\odot{}$) and high-mass ($\gtrsim{}10\,{}M_\odot{}$) companions  generate transient (with a bright phase of only a few days) and persistent bright X-ray sources, respectively.
As $10\,{}M_\odot{}$ stars have a lifetime of $\sim{}30-40\,{}$ Myr, only IMBHs hosted in sufficiently young star clusters can generate persistent X-ray sources. Then, IMBHs hosted in clusters born before the dynamical encounter with the intruder (i.e. more than 100 Myr ago) can produce only transient sources.

We estimated that, out of 100 IMBHs which were present before the dynamical encounter with the intruder galaxy, only $\lesssim{}2-3$ are expected to undergo mass transfer from low-mass companions at present, producing a comparable number of transient X-ray sources. As observations show that at least 4 X-ray sources in the Cartwheel ring are persistent over a time-scale of 6 months (Wolter et al. 2006), we conclude that pre-encounter formed IMBH binaries are not sufficient to explain the data.

We considered the possibility that pre-encounter IMBHs capture massive stars produced after the encounter with the intruder. In this case, under overoptimistic assumptions, 100 pre-encounter IMBHs can produce $\lesssim{}1$ X-ray source, either persistent or not.

Finally, we hypothesized that very young ($<40$ Myr) star clusters, formed after the encounter, generate IMBHs at their centre. Under this hypothesis, $500-1000$ IMBHs are required to produce $\sim{}15-30$ bright ($10^{36}-10^{41}$ erg s$^{-1}$) X-ray sources, some of them persistent and  some transient
. This scenario might account for the $\sim{}17$  observed X-ray sources in the Cartwheel ring. It is also in agreement with the fact that many ULXs observed in Cartwheel are associated with bright $H_\alpha{}$ spots, i.e. active star-forming regions (Gao et al. 2003).


The birth of $\sim{}1000$ IMBHs (each one of 100 $M_\odot{}$) in $\sim{}$40 Myr  implies an IMBH formation rate of $2.5\times{}10^{-3}\,{}M_\odot{}\,{}{\rm yr}^{-1}$, that is a factor of $\sim{}10^4$ lower than the SF rate. This rate is acceptable for runaway collapse scenarios, as Portegies Zwart \& McMillan (2002) show that  $\sim{}$0.1 per cent of the mass of the parent young cluster merges to form the IMBH.

 However, we stress that only the runaway collapse scenario, under extreme assumptions, can explain the formation of such a huge number of IMBHs in the disc. Thus, our simulations suggest that IMBHs can hardly account for all the ULXs observed in Cartwheel.

On the other hand, it is  possible that only the few brightest sources in Cartwheel  are powered by IMBHs, while the other ones are either beamed HMXBs, super-Eddington accreting stellar mass BHs  or a blending of multiple fainter sources. For example, $\sim{}30$ IMBHs are expected to form in the ring, in order to produce only 1  very bright ULX, such as the source N.10 in Cartwheel.

These results agree with the semi-analytical model by King (2004), who showed that IMBHs cannot explain all the ULXs in Cartwheel. However, King (2004) concludes that $>3\times{}10^4$ IMBHs are required to produce the observed number of ULXs, $\sim{}30-60$ times more than in our analysis. This apparent discrepancy is due to the fact that  King (2004) assumes that the IMBHs power only transient sources (Kalogera et al. 2004), and thus he has to introduce a $\sim{}10^{-2}$ duty-cycle. However, Patruno et al. (2005) showed that IMBHs accreting from young massive stars ($\gtrsim10\,{}M_\odot{}$) produce non-transient sources, increasing the expected duty-cycle.


In conclusion, new $Chandra$ and $XMM-Newton$ observations of Cartwheel could partially  solve the mystery of Cartwheel X-ray sources, investigating which sources are transient, which variable, and which persistent. Deeper observations are also needed  to resolve possible blended sources. 

In the future, it would be interesting to search whether other ring galaxies 
host  as many bright X-ray sources as Cartwheel and whether these sources are similarly concentrated in the outer ring.


\section*{Acknowledgments}
The authors thank J.~Stadel, P.~Englmaier and D.~Potter for technical support, and acknowledge the two referees for the critical reading of the manuscript and for their helpful comments. We also thank A.~Wolter and G.~Trinchieri for their useful comments. MM acknowledges support from the Swiss
National Science Foundation, project number 200020-109581/1
(Computational Cosmology \&{} Astrophysics).

{}

\begin{appendix}
\section{The formation of spokes}
Well-developed spokes appear in our simulations, especially in runs A1, A2 and A3 (see Fig.~\ref{fig:fig2}). These spokes are composed mainly by stars and are quite gas-poor, in agreement with observations (Higdon 1995, 1996; Mayya et al. 2005), but at odd with previous simulations.
In fact, Hernquist \& Weil (1993) and Mihos \& Hernquist (1994) predict very gas-rich spokes, born from fragmentation of the inner gas ring. Instead, we do not observe any significant fragmentation of the gas ring during the entire simulation ($t=1$ Gyr). Artificial fragmentation can occur in simulations, due to low resolution. As previous simulations (Hernquist \& Weil 1993; Mihos \& Hernquist 1994) have a factor of $\sim{}10$ less particles than ours, we ran some low resolution check, and we observed an analogous fragmentation of the gas ring.

\begin{figure}
\center{{
\epsfig{figure=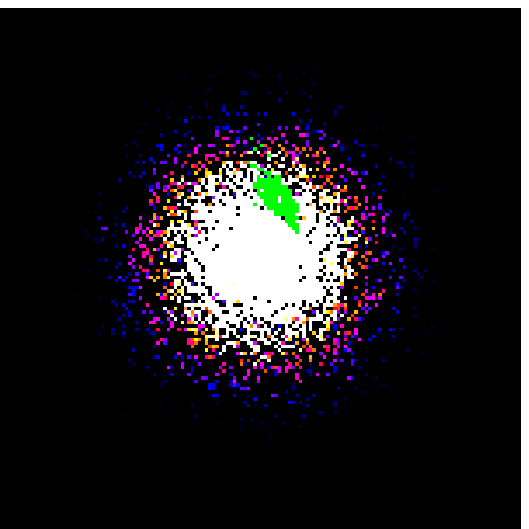,height=4cm}
\epsfig{figure=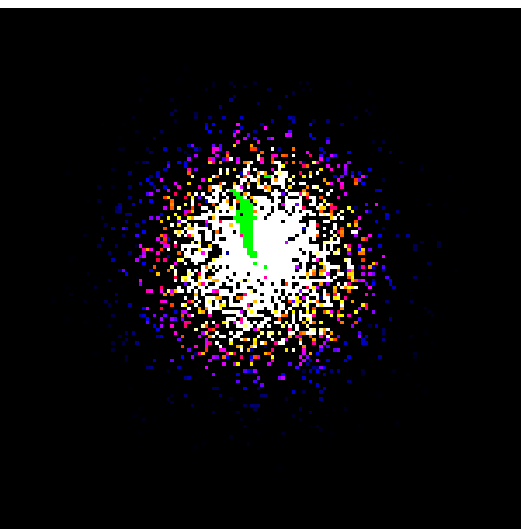,height=4cm}
\epsfig{figure=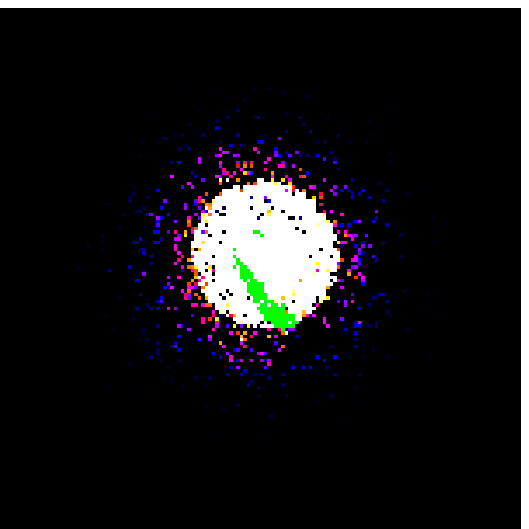,height=4cm}
\epsfig{figure=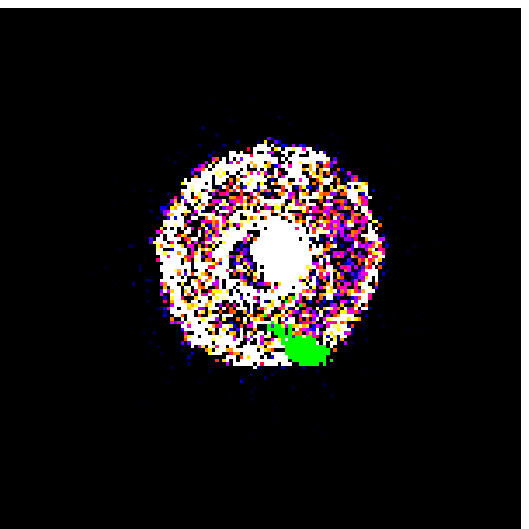,height=4cm}
\epsfig{figure=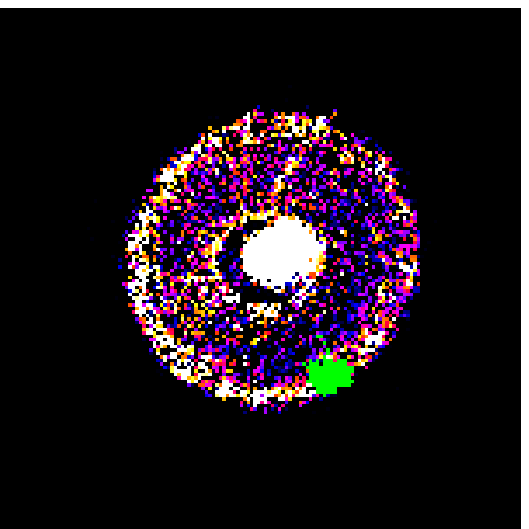,height=4cm}
\epsfig{figure=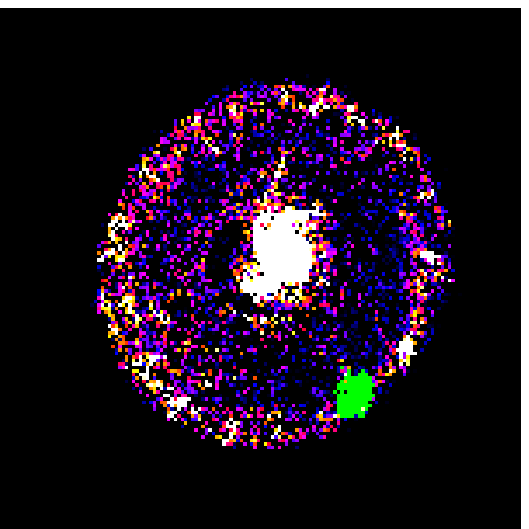,height=4cm}
\epsfig{figure=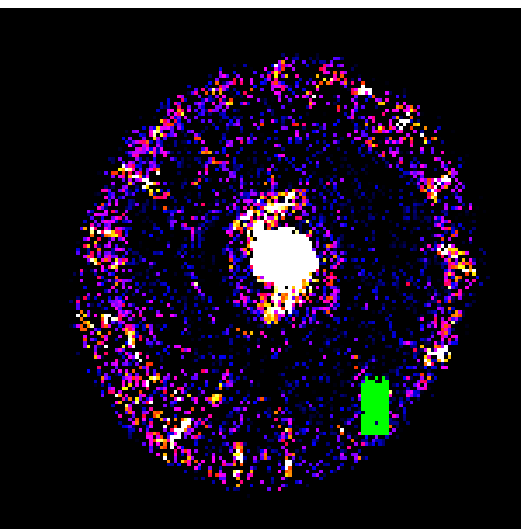,height=4cm}
\epsfig{figure=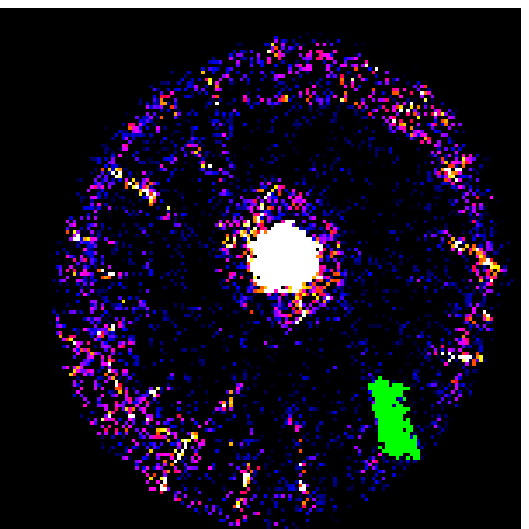,height=4cm}
}}
\caption{\label{fig:fig12} 
Run A2. From top to bottom and from left to right: face-on plot of the stellar density at $t=20$, 40, 60, 80, 100, 120, 140, 160 Myr. The color coding is the same as in Fig.~\ref{fig:fig6}. The filled circles (green in the online version) are marked stellar particles belonging to one of the well-developed spokes. Each frame measures 100 kpc per edge.} 
\end{figure}

But what is the physical mechanism which leads to the formation of stellar rather than gaseous spokes? In Fig.~\ref{fig:fig12} we selected some of the stellar particles composing one of the best developed spokes  at $t=140$ Myr (in run A2), and we followed them backward in time. These particles resulted to be close to each other since the beginning of the simulation. During the crucial phase of the collision (i.e. between $t=$40 and $t=60$ Myr) they happen to be close to both the nuclei of the two interacting galaxies and they are accelerated and ejected in the same direction. 

This process looks like a 'sling-shot' interaction, similar to what happens during three-body encounters: the intruder interacts with the nucleus of the Cartwheel and ejects bound stars in different directions. Stars which before the interaction were tightly close to each other (having approximately the same velocity) are ejected more or less in the same direction and form the spokes.

We repeated this procedure, marking bunches of stars from several different spokes and tracking them backward in time, and we always found the same dynamics: particles which were close at the beginning are 'ejected' in similar orbits. 

 These 'ejected' stars  end up in the stellar ring, which  appears quite smooth during the first stages of its life ($t<120$ Myr). At $t\sim{}120$ Myr local gravitational instabilities start developing in the stellar ring: quasi-spherical clumps of stars form all along the ring.

The particles in each clump have a range of velocities such that each clump
becomes more and more stretched ($t\gtrsim{}140$ Myr). In fact, some particles have smaller radial velocity or have already reached the turn around, and start falling back to the centre. Other particles have still enough radial velocity to continue their expansion (up to $t\gtrsim{}200$ Myr). These stretched clumps reproduce quite well the observed spokes in Cartwheel.

Also the profile of the expansion velocity in Cartwheel (Higdon 1996) supports this idea: the gas beyond 10-15 kpc is still expanding, while the one within this distance is already falling back.


The hypothesis that gravitational instabilities in the stellar ring are the source of spokes is also supported by the fact that the spokes tend to disappear if we reduce the mass of the disc, or increase the concentration of the halo, in both cases increasing the stability of the disc. For instance, in runs B the mass of the stellar disc is one half than in runs A and the spokes are much less evident. Similarly, if we increase by $\gtrsim{}10$ per cent the dark matter mass within $r_{s}$, the formation of spokes is dramatically quenched even in a run identical to A1 for all the other characteristics.

This dependence of spoke formation on the disc mass and on the central halo mass puts strong constraints on the disc and halo parameters of Cartwheel, i.e. it suggests that Cartwheel should be close to the maximum-disc model (van Albada \& Sancisi 1986). This result agrees with the observational properties of Cartwheel, which indicate a total mass of $\sim{}6\times{}10^{11}\,{}M_\odot{}$ (Higdon 1996) and a disc mass of $\sim{}6\times{}10^{10}\,{}M_\odot{}$ (Marcum et al. 1992).

In conclusion, only for a small range of initial conditions we obtain well-developed spokes (such as in runs A). This result agrees with observations, as a small fraction (less than $\sim{}10$ per cent) of ring galaxies shows detectable spokes. It would be interesting to compare the stellar disc mass and the total dynamical mass of other ring galaxies. If there is a correlation between the presence of the spokes, the stellar disc mass and the mass of the halo, this would support our hypothesis.


\end{appendix}

\end{document}